%% file: main.tex
\documentclass[times,doublespace]{stvrauth}

\usepackage{moreverb}

\usepackage[colorlinks,bookmarksopen,bookmarksnumbered,citecolor=red,urlcolor=red]{hyperref}

\newcommand\BibTeX{{\rmfamily B\kern-.05em \textsc{i\kern-.025em b}\kern-.08em
T\kern-.1667em\lower.7ex\hbox{E}\kern-.125emX}}

\def\volumeyear{2017}

\usepackage{amssymb}
\usepackage{amsmath}
\DeclareMathOperator*{\argmax}{arg\,max}
\usepackage{array}
\newtheorem{definition}{Definition}
\usepackage{centernot}
\usepackage{mathtools}
\usepackage{tabularx}
\usepackage{xcolor}
\usepackage{booktabs}
\usepackage{verbatim}
\usepackage{algorithm}
\usepackage[noend]{algpseudocode}
\usepackage{subfig}
\usepackage{framed}

\newcommand{\etal}{\textit{et al.}}

\newcommand{\rev}[1]{{\color{black} #1}}
\newcommand*{\rom}[1]{\textup{\uppercase\expandafter{\romannumeral#1}}}
\newcommand{\specialcell}[2][l]{\begin{tabular}[#1]{@{}l@{}}#2\end{tabular}}

\begin{document}

\runningheads{D. Shin et al.}{Empirical Evaluation of Mutation-based Test Prioritization Techniques}

\title{Empirical Evaluation of Mutation-based \\Test Case Prioritization Techniques}

\author{Donghwan Shin\affil{1}\corrauth, Shin Yoo\affil{1}, Mike Papadakis\affil{2}, Doo-Hwan Bae\affil{1}}

\address{
\affilnum{1}KAIST, 291 Daehak-ro Yuseong-gu, Daejeon, Republic of Korea\break
\affilnum{2}University of Luxembourg, Luxembourg
}

\corraddr{KAIST, 291 Daehak-ro Yuseong-gu, Daejeon, Republic of Korea. E-mail: donghwan@se.kaist.ac.kr}

\begin{abstract}

We propose a new test case prioritization technique that combines both mutation-based and diversity-based approaches. Our diversity-aware mutation-based technique relies on the notion of mutant distinguishment, which aims to distinguish one mutant's behavior from another, rather than from the original program. We empirically investigate the relative cost and effectiveness of the mutation-based prioritization techniques (i.e., using both the traditional mutant kill and the proposed mutant distinguishment) with 352 real faults and 553,477 developer-written test cases. The empirical evaluation considers both the traditional and the diversity-aware mutation criteria in various settings: single-objective greedy, hybrid, and multi-objective optimization. The results show that there is no single dominant technique across all the studied faults. To this end, \rev{we we show when and the reason why each one of the mutation-based prioritization criteria performs poorly, using a graphical model called Mutant Distinguishment Graph (MDG) that demonstrates the distribution of the fault detecting test cases with respect to mutant kills and distinguishment.}
\end{abstract}


\maketitle

\input{intro}
\input{background}
\input{techniques}

\input{expr}
\input{result}
\input{relatedWork}
\input{conclusion}

\bibliographystyle{wileyj}
\bibliography{mutation.bib}
\end{document}

%% file: intro.tex
\section{Introduction}\label{sec:intro}

Test case prioritization aims at ordering regression test suites so that testing meets its goals as early as possible. This means that stopping the test process at any arbitrary point (in time), test effectiveness is optimal (with respect to the used time budget)~\cite{yoo2012regression}. To achieve this goal, test prioritisation needs to `predict' which test cases will detect faults. This prediction is usually performed based on surrogates, like test coverage~\cite{rothermel2001prioritizing} or test case diversity~\cite{henard2014bypassing, henard2016comparing, feldt2016test}.  

Although test prioritization techniques have been extensively studied in the literature~\cite{henard2016comparing}, most of them rely on the use of various types of structural coverage~\cite{yoo2012regression}. Little attention has been paid to advanced test elements like mutants (i.e., artificial faults). We believe the mutation based criteria call for further attention, given that mutants have been shown to be effective at revealing faults~\cite{TitcheuPTH17} and that mutant killing (i.e., detecting the deference between a mutant and its original program) ratios are similar with the fault detection ratios~\cite{rothermel2001prioritizing, andrews2006using}. Yet very few approaches study the mutation-based test case prioritization and none evaluates them with real-world applications and faults.
 
Recent advances in test case prioritization focus on identifying and promoting the diversity of the selected test cases~\cite{henard2014bypassing, henard2016comparing, feldt2016test}, rather than maximising the coverage. This trend can provide several advantages, especially in cases where there is no source code availability~\cite{henard2016comparing}. Therefore, the combination of mutation-based and diversity-based approaches could provide substantial benefits by increasing early fault detection. Investigating such a combination is the primary aim of this study.

In this paper we propose and empirically investigate a new diversity-aware mutation-based test case prioritization technique. The technique relies on the diversity-aware mutation adequacy criterion, which is recently proposed by Shin et al.~\cite{shin2016diversity,shin2017theoretical}. The diversity-aware criterion aim is to distinguish the behaviour of every mutant from the behaviour of all other mutants (and the original program version). In contrast the traditional mutation adequacy criterion aims at distinguishing only the behaviour of the mutants from that of the original program. According to Shin et al., distinguishing mutants improves the fault detection capabilities of mutation testing. Therefore, our diversity-aware mutation-based prioritization gives higher priority to those test cases that help distinguishing all mutants as early as possible.

Our study investigates the relative cost and effectiveness of two mutation-based prioritization techniques, i.e., one using traditional mutant kill and another using distinguishement, with real-world applications and faults. For this, we use 352 real faults and 553,477 developer-written test cases in the \texttt{Defects4J} data set~\cite{just2014defects4j}. The empirical evaluation considers both the traditional kill-only and the proposed diversity-aware mutation-based prioritization criteria in various settings: single-objective greedy, single-objective hybrid, as well as multi-objective prioritization that seeks to prioritize using both criteria simultaneously. We find that there is no single technique that we could characterize as the dominant one. To this end, \rev{we provide a graphical model called Mutant Distinguishment Graph (MDG) to help us understand how a set of test cases that kills and distinguishes mutants related with fault detection. This visualisation scheme demonstrates why and when each one of the mutation-based prioritization criteria performs poorly.}

Overall, the technical contributions of this paper can be summarized as follows:
\begin{itemize}
\item We present a large empirical study that investigates the relative cost and effectiveness of mutation-based prioritization techniques with real faults.
\item We investigate two different mutation-based test prioritization techniques under both single (greedy and hybrid) and multi-objective prioritization schemes.
\item We investigate and identify the reasons behind the differences between the traditional kill-only mutation and distinguish mutation prioritization schemes, using intuitive graphical models.
\end{itemize}

The rest of this paper is organized as follows. Section~\ref{sec:background} provides background for mutation adequacy criteria and test case prioritization. Section~\ref{sec:mutationPrt} explains the mutation-based test case prioritization techniques that are studied in this paper. Section~\ref{sec:expr} explains our experimental settings including research questions, measures and variables, subject faults, test, and mutants. The results from the empirical evaluation are given in Section~\ref{sec:results}, together with the threats to validity. Section~\ref{sec:related} presents the related work, and Section~\ref{sec:conclusion} concludes this paper.

%% file: background.tex
\section{Background}\label{sec:background}

\subsection{Mutation Adequacy Criteria}\label{sec:mutationAdequacy}
In the late 1970s, DeMillo et al.~\cite{demillo1978hints} proposed the mutation adequacy criterion as a way to assess the quality of test suites. The criterion focuses on the differences between the original program version and its mutant versions (i.e., artificially mutate programs) in the program outputs to measure the mutant kills. This technique is relies on the idea that tests capable of distinguishing the behavior of mutants from those of the original programs are also capable of revealing faults. This idea was recently extended by Shin et al.~\cite{shin2017theoretical}, who proposed distinguishing the behavior of mutants among themselves (in addition to the original program). This forms a diversity-aware mutation adequacy criterion that caters for the diversity of behaviors introduced by the mutants. 

To be precise, we formally represent and discuss the mutation adequacy criteria using the essential elements of an existing formal framework (for the mutation-based testing methods)~\cite{shin:2016theoretical}. Let $P$ be a set of programs which includes the program under test. There are an original program $p_o\in P$ and a mutant \rev{$m\in M\subset P$} generated from $p_o$. For a test case $t$ in a set of test cases $T$, if the behaviors of $p_o$ and $m$ are different for $t$, it is said that $t$ \textit{kills} $m$. Note that the notion of behavioral difference is an abstract concept. It is formalized by a testing factor, called a test differentiator, which is defined as follows:
\begin{definition}[Test Differentiator]
A \textit{test differentiator} $d: T\times P\times P \to \{0,1\}$ is a function,\footnote{This function-style definition is replaceable by a predicate-style definition, such as $d\subseteq T\times P\times P$.} such that
\begin{displaymath}
d(t,p_x,p_y) = 
\begin{cases}
1\text{ }(true), & \text{if the behaviors of $p_x$ and $p_y$ are \textit{different} for $t$} \\ 
0\text{ }(false), & \text{otherwise}
\end{cases}
\end{displaymath}
for all test cases $t\in T$ and programs $p_x,p_y\in P$.
\end{definition}
By definition, a test differentiator concisely represents whether the behaviors of $p_x\in P$ and $p_y\in P$ are different for $t$. In addition to a differentiator which formalizes the difference of two programs for a single test, it will be helpful to consider whether the two programs are different for a set of tests. A \textit{d-vector} is defined to represent such difference of the programs as follows:
\begin{definition}[d-vector]\label{def:d-vector} 
A \textit{d-vector} $\mathbf{d}:T^n\times P\times P \to \{0,1\}^n$ is an $n$-dimensional vector, such that
\begin{displaymath}
\mathbf{d}(TS,p_x,p_y) = \langle d(t_1,p_x,p_y), ..., d(t_n,p_x,p_y) \rangle
\end{displaymath}
for all $TS=\{t_1,\cdots,t_n\} \subseteq T^n$, $d\in D$, and $p_x,p_y\in P$.
\end{definition}
In other words, a differentiator $d$ returns Boolean value (i.e., 0 or 1) from a single test, whereas a d-vector $\mathbf{d}$ returns $n$-dimensional Boolean vector from $n$ test cases. \rev{Note that a test suite $TS$ is used to denote the order of test cases in a the test suite, while $T$ denotes a set of tests without any particular order of test cases.}

Using the test differentiator and d-vector, we define the notion of mutant kill as follows:
\begin{definition}[Mutant Kill]
A mutant $m$ generated from an original program $p_o$ is \emph{killed} by a test case $t$ when the following condition holds:
\begin{displaymath}
d(t,p_o,m) \neq 0.
\end{displaymath}
Similarly, $m$ generated from $p_o$ is \emph{killed} by a test suite $TS$ when the following condition holds:
\begin{displaymath}
\mathbf{d}(TS,p_o,m) \neq \mathbf{0}.
\end{displaymath}
\end{definition}

\begin{figure}
	\centering
	\includegraphics[width=0.7\linewidth]{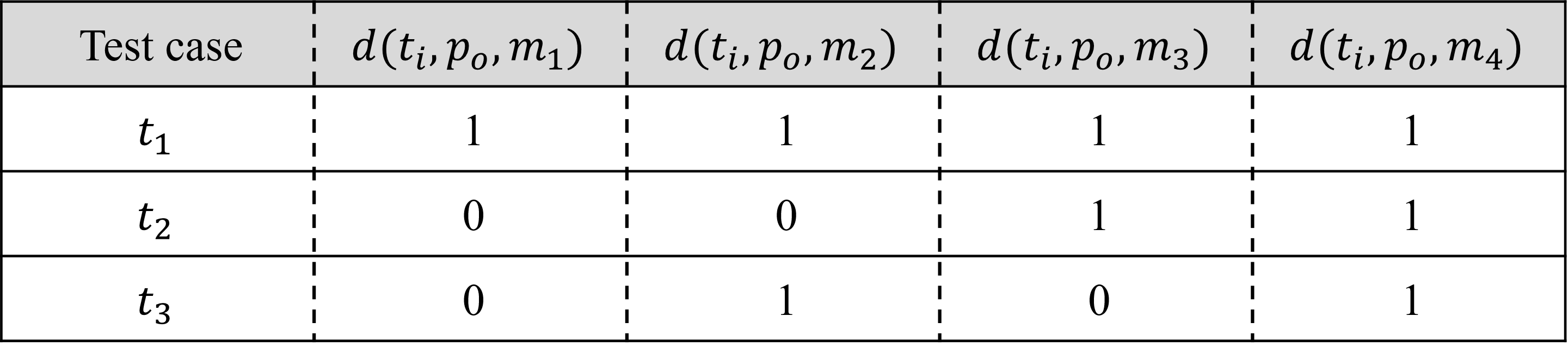}
	\caption{A working example for mutation adequacy criteria. The table represents whether a test case kills a mutant. For example, $d(t_1, p_o, m_1)$ is 1 which means that $t_1$ kills $m_1$.}
	\label{fig:example}
\end{figure}

Based on the notion of mutant kill, the traditional mutation adequacy criterion is defined as follows:
\begin{definition}[Traditional Mutation Adequacy Criterion]\label{def:k}
For a set of mutants $M$ generated from an original program $p_o$, a test suite $TS$ is \emph{mutation-adequate} when the following condition holds:
\begin{displaymath}
\forall m\in M, \mathbf{d}(TS,p_o,m) \neq \mathbf{0}.
\end{displaymath}
\end{definition}
This definition means that a test suite $TS$ is mutation-adequate if all mutants $m\in M$ are killed by at least one test case $t\in TS$. Figure~\ref{fig:example} shows the working example for mutation adequacy criteria. There are four different mutants and three different test cases. Each of the values represents whether a test case kills a mutant. In the working example, a test suite $TS_1= \{t_1 \}$ is adequate to the traditional mutation adequacy criterion because all the four mutants are killed by $t_1$. \rev{This working example is an abstraction of what commonly happens in practice. For example, it is very common that several test cases, like $t_1$, ``thinly" inspect a large part of a program, while the others, like $t_2$ or $t_3$, ``deeply" inspect a specific part of the program.}

Note that the traditional mutation adequacy focuses on the difference between a mutant and its original program. To formalize the diversity of mutants in terms of test cases, the notion of mutant distinguishment is defined as follows:
\begin{definition}[Mutant Distinguishment]
Two mutants $m_x$ and $m_y$ generated from an original program $p_o$ are \emph{distinguished} by a test case $t$ when the following condition holds:
\begin{displaymath}
d(t,p_o,m_x) \neq d(t,p_o,m_y).
\end{displaymath}
Similarly, $m_x$ and $m_y$ generated from $p_o$ are \emph{distinguished} by a test suite $TS$ when the following condition holds:
\begin{displaymath}
\mathbf{d}(TS,p_o,m_x) \neq \mathbf{d}(TS,p_o,m_y).
\end{displaymath}
\end{definition}

We now introduce the diversity-aware mutation adequacy criterion, called the \emph{distinguishing mutation adequacy criterion}, based on the mutant distinguishment as follows:
\begin{definition}[Distinguishing Mutation Adequacy Criterion]\label{def:newMutationAdequacy} 
For a set of mutants $M$ generated from an original program $p_o$, a test suite $TS$ is \textit{distinguishing mutation-adequate} when the following condition holds:
\begin{displaymath}
\forall m_x,m_y\in M', \mathbf{d}(TS,p_o,m_x) \neq \mathbf{d}(TS,p_o,m_y)
\end{displaymath}
where $m_x \neq m_y$ and $M' = M \cup \{p_o\}$.
\end{definition}
In other words, a test suite $TS$ is distinguishing mutation-adequate if all possible pair of different mutants $m_x$ and $m_y$ in $M'$ are distinguished by $TS$. In the working example, $TS_3=\{t_1,t_2,t_3\}$ distinguishes all mutants in $M'=\{p_o,m_1,\cdots,m_4\}$ because all five mutants (i.e., $p_o,m_1,m_2,m_3,m_4$) have unique d-vectors for $TS_3$. 

For the sake of simplicity, let $d$-criterion hereafter refer to the distinguishing mutation adequacy criterion (i.e., diversity-aware) and, similarly, $k$-criterion to the traditional mutation adequacy criterion (i.e., kill-only).

\rev{By definition, the $d$-criterion \emph{subsumes} the $k$-criterion: for a set of mutants $M$ generated from an original program $p_o$, if a test suite $TS$ is adequate to the $d$-criterion, it is guaranteed that $TS$ is adequate to the $k$-criterion as well. In other words, the $d$-criterion is stronger than the $k$-criterion. For more information, please refer the recent study of Shin et al.~\cite{shin2017theoretical}.}

\subsection{Test Case Prioritization}\label{sec:background-prt}
\begin{figure}
	\centering
	\includegraphics[width=0.8\linewidth]{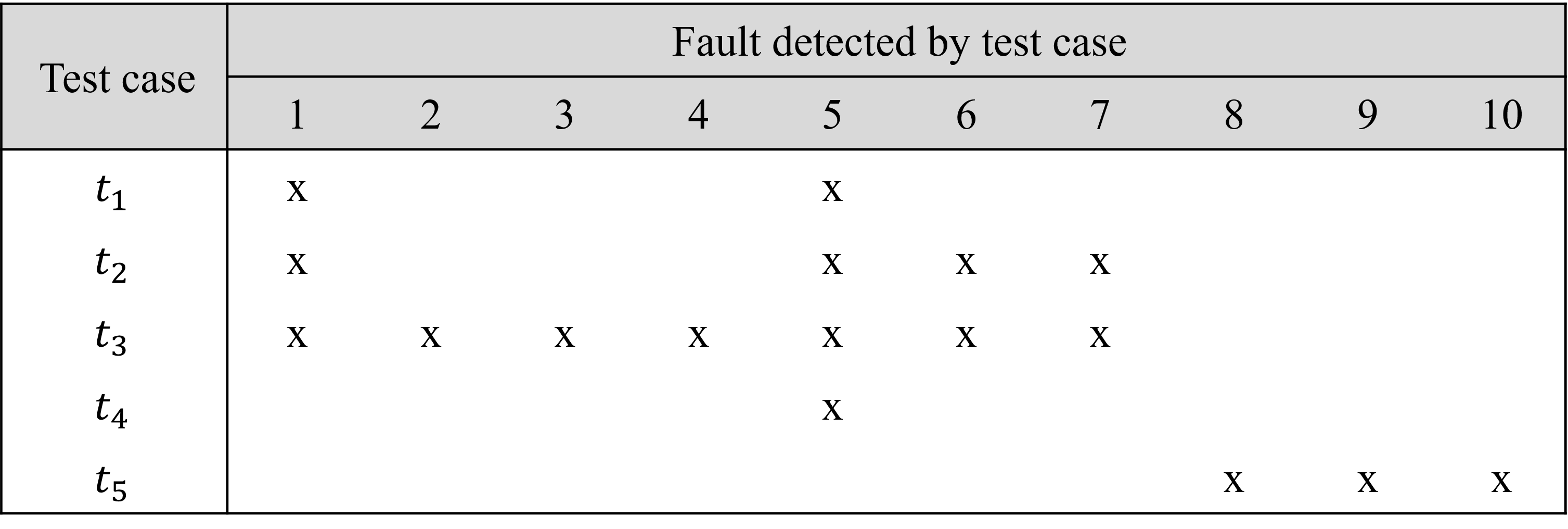}
	\caption{\rev{Example test suite with fault detection information, taken from Rothermel \etal~\cite{rothermel2001prioritizing}. Executing $t_3$ followed by $t_5$ is clearly beneficial in early fault detection.}}
	\label{fig:prt-example}
\end{figure}

\rev{Regression testing is performed when changes are made to existing software; the purpose of regression testing is to provide confidence that the newly introduced changes do not obstruct the behaviors of the existing, unchanged part of the software~\cite{yoo2012regression}. In regression testing, test case prioritization finds an ordering of test cases that maximizes a desirable property, such as the rate of fault detection. To see the benefit of using test case prioritization, consider the test suite with fault detection information in Figure~\ref{fig:prt-example}. If a tester want to detect faults as early as possible, it is clearly beneficial to execute $t_3$ first, followed by $t_5$.}

Rothermel \etal~\cite{rothermel2001prioritizing} formally define the test case prioritization problem as follows:

\begin{definition}[Test Case Prioritization Problem]\label{def:prt}
Given: A test suite, $TS$, the set of permutations of $TS$, $\Pi$, and an objective function from $\Pi$ to real numbers, $f: \Pi \rightarrow \mathbb{R}$.\\
Problem: Find a permutation $\pi\in \Pi$ such that $\forall \pi' \in\Pi, (\pi' \neq \pi) \wedge (f(\pi) \ge f(\pi'))$.
\end{definition}

In this definition, $\Pi$ represents all possible orderings of the given test cases in $TS$, and $f$ represents an objective function that calculates an award value for an ordering $\pi\in\Pi$. \rev{For the example in Figure~\ref{fig:prt-example}, an ordering $\pi_x = \langle t_3, t_5, t_1, t_2, t_4\rangle$ is better than another ordering $\pi_y=\langle t_1, t_2, t_3, t_4, t_5 \rangle$, since $\pi_x$ detects the fault earlier than $\pi_y$.}

\rev{The main usage scenario of the prioritisation techniques is to be used for the test of the program changes made on subsequent program versions. Recent research \cite{henard2016comparing} has shown that the effectiveness degradation of the prioritization techniques over subsequent program versions is small and that taking into account the code changes performed on a subsequent version does not provide any important information \cite{LuLCZHZ016}. Therefore, testers need to obtain the required information at a specific point in time (prioritization time) and then use it to prioritize and order the relevant test suites in the subsequent program versions. 

At prioritization time, we need to consider a surrogate for the fault detection based on the historical information of the test cases instead of re-executing them, hoping that early maximization of the surrogate will result in the early maximization of the fault detection. Therefore, while the goal of test case prioritization remains the early maximization of the fault detection, it actually aims the early maximization of the chosen surrogate. Naturally, the test case prioritization techniques vary depending on the chosen surrogate.

The structural coverage information, such as statement coverage, of test cases is one of the widely-used surrogate in test case prioritization~\cite{rothermel2001prioritizing,elbaum2002test,zhang2013bridging}. For example, the statement-total approach prioritizes test cases according to the number of statements covered by individual test cases. In other words, a test case covering more statements has higher priority. Similarly, the statement-additional approach prioritizes test cases according to the additional number of statements covered by individual test cases.

Mutants are also used as another surrogate for test case prioritization~\cite{rothermel2001prioritizing,do2006use,lou2015mutation}. Instead of using the structural coverage of individual test cases, the mutant kill of individual test cases is utilized. For example, Rothermel \etal~\cite{rothermel2001prioritizing} consider the Fault Exposing Potential (FEP)-total approach that prioritizes test cases according to the number of mutants killed by individual test cases. Similarly, the FEP-additional approach prioritizes test cases according to the additional number of mutants killed by individual test cases. Note that, to kill a mutant, a test case not only needs to cover the location of mutation but also to execute the mutated part~\cite{yoo2012regression}. It means the mutation-based approaches can be constructed at least as strong as coverage-based approaches. 

In this paper, we focus on the mutation-based test case prioritization, using the two mutation-based adequacy criteria (i.e., kill and distinguish), while we use the coverage-based and random approaches as baselines.}

\subsection{Multi-Objective Test Case Prioritization}\label{sec:multiObj}
The essence of the multi-objective optimization is the notion of Pareto optimality. Given multiple objectives, an ordering of test cases is said to be \emph{non-dominated} if none of the objectives can be improved in value without degrading the other objective values. Otherwise, an ordering of test cases is said to be \emph{dominated} by another ordering that has at least one higher objective value without decreasing any other objective values. Formally, let $O$ be the number of different objectives. For $i\in\{1,2,\cdots,O\}$, each objective function is represented as $f_i: \Pi\rightarrow\mathbf{R}$. An ordering $\pi_x$ is said to dominate another ordering $\pi_y$ if and only if the following is satisfied:
\begin{displaymath}
(\forall i\in \{1,2,\cdots,O\}, f_i(\pi_x) \ge f_i(\pi_y))
\wedge (\exists i\in \{1,2,\cdots,O\}, f_i(\pi_x) > f_i(\pi_y))
\end{displaymath}

When evolutionary algorithms are applied to multi-objective optimization, they produce a set of orderings that are not dominated by each other. Such a set is called a Pareto front. The number of orderings in a Pareto front is determined by the number of population in the evolutionary algorithms. For example, the Non-Dominated Sorting Genetic Algorithm \rom{2} (NSGA-\rom{2})~\cite{deb2002fast}, one of the most widely studied multi-objective evolutionary algorithm, generates \rev{$K$} number of Pareto optimal solutions in a Pareto front\rev{, where $K$ is the predefined population size.}

%% file: techniques.tex
\section{Mutation-based Test Case Prioritization Techniques}\label{sec:mutationPrt}

In this paper, we consider six different test case prioritization techniques as described in Table~\ref{table:techniques}. The first column represents the mnemonic for each technique that will be used throughout this paper. The second column represents the prioritization objective of each technique. The letters for the mnemonic are capitalized. The third column represents the tie-breaking rule when there are multiple candidate test cases (for greedy and hybrid) or orderings (for multi-objective optimization) satisfying the same level of the objective(s). The last column summarizes each technique. \rev{Additional details regarding the techniques listed in Table~\ref{table:techniques} can be found in the following subsections.}

\begin{table}[htb]
\centering
\caption{Summary of mutation-based test case prioritization techniques}
\label{table:techniques}
\begin{tabularx}{\textwidth}{lllX}
\toprule
Mnemonic & Objective & Tiebreaker & Description \\
\midrule
\midrule
GRK & \specialcell[t]{GReedy,\\Kill}	& random	& iteratively select a test case that maximizes the number of additionally killed mutants \\
GRD & \specialcell[t]{GReedy,\\Distinguish}	& random	& iteratively select a test case that maximizes the number of additionally distinguished mutants \\
HYB-$w$	& \specialcell[t]{HYBrid,\\$w$eight}	& random	& iteratively select a test case that maximizes the weighted sum of the number of additionally killed mutants and additionally distinguished mutants \\
\midrule
MOK		& \specialcell[t]{Multi-Objective,\\kill \&\\distinguish}	& kill & optimize an ordering of test cases to both kills and distinguishes mutants as early as possible, and select one of the Pareto optimal orderings that kills mutants as early as possible \\
MOD		& \specialcell[t]{Multi-Objective,\\kill \&\\distinguish}	& distinguish & optimize an ordering of test cases to both kills and distinguishes mutants as early as possible, and select one of the Pareto optimal orderings that distinguishes mutants as early as possible \\
\midrule
RND & RaNDom	& random	& randomized ordering \\
\rev{SCV} & \specialcell[t]{Statement\\CoVerage} & random & iteratively select a test case that maximizes the number of additionally covered statements \\
\bottomrule
\end{tabularx}
\end{table}

\rev{\subsection{Greedy and Hybrid Techniques}}

\rev{We first describe the single-objective greedy techniques: GRK, GRD, and HYB. Algorithmically, these techniques are in essence instances of additional greedy algorithms~\cite{li2007search}. The additional greedy test case prioritization technique iteratively selects a test case that maximizes the additional achievement of the objective at a time. Note that the hybrid technique is also an instance of the single-objective additional greedy because its only objective is the form of the weighted sum of GRK and GRD.}

\paragraph{GRK and GRD:}
Based on the $k$-criterion, GRK iteratively selects a test case that maximizes the number of additionally killed mutants. If there are multiple test cases that additionally kills the same number of mutants, one of them is randomly selected. Formally, let $\kappa(t)$ be the number of additional mutants killed by a test case $t$. GRK iteratively selects $t$ in a  test suite $TS$ that satisfies $\argmax_{t\in TS}(\kappa(t))$. Similarly, GRD iteratively selects a test case that maximizes the number of additionally distinguished mutants, based on the $d$-criterion. If there are multiple test cases that additionally distinguishes the same number of mutants, one of them is randomly selected. Formally, let $\delta(t)$ be the number of additional mutants distinguished by a test case $t$. GRD iteratively selects $t$ in a test suite $TS$ that satisfies $\argmax_{t\in TS}(\delta(t))$.

Semantically, GRK distinguishes mutants from its original program as early as possible, whereas GRD distinguishes all mutants from each other as early as possible. In other words, GRK is essentially based on the concept of intensification, whereas GRD is essentially based on the concept of diversification. Such difference may lead the effectiveness difference between GRK and GRD in prioritization. Section~\ref{sec:discussion} discusses this issue in more detail.

\rev{As we explained in Section~\ref{sec:background-prt}, GRK is another name of FEP-additional used by Rothermel et al.~\cite{rothermel2001prioritizing}. Since they already report that FEP-additional is more effective than FEP-total, we only consider the additional approaches for the our greedy techniques.}

\paragraph{HYB-$w$:}
This hybrid prioritization technique is a weighted sum of GRK and GRD. It iteratively selects a test case that maximizes the number of the weighted sum of additionally killed mutants and additionally distinguished mutants. Formally, for a weight factor $w\in [0, 1]$, HYB-$w$ iteratively selects a test case $t$ in a test suite $TS$ that $\argmax_{t\in TS}(w\times \kappa(t) + (1-w)\times \delta(t))$. By definition, $w=1$ refers the GRK technique and $w=0$ refers the GRD technique.


\rev{\subsection{Multi-Objective Optimization Techniques}}
\rev{Unlike the greedy techniques, which iteratively select a test case that suits its objective in a given situation, a multi-objective prioritization technique optimizes an ordering of test cases as a whole to both kill and distinguish mutants as early as possible.}

\begin{figure}
	\centering
	\includegraphics[width=0.6\linewidth]{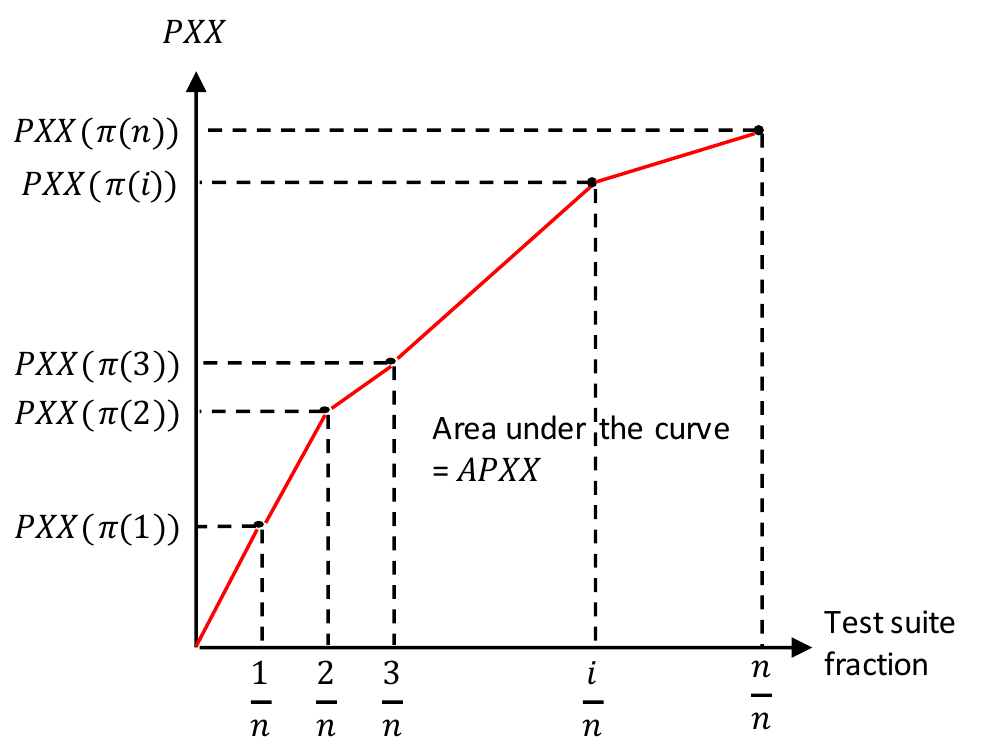}
	\caption{The concept of APXX. The area under the curve is the APXX value.}
	\label{fig:apxx}
\end{figure}

\paragraph{MOK and MOD:} 
\rev{To represent the two mutation-based objectives (i.e., \emph{kill} mutants as early as possible and \emph{distinguish} mutants as early as possible) as two measurable functions (i.e., fitness functions in an evolutionary algorithm),} we define metrics called APMK (Average Percentage of Mutants Killed) and APMD (Average Percentage of Mutants Distinguished), respectively. The core of these metrics are in APFD (Average Percentage of Faults Detected)~\cite{elbaum2002test} that is the most commonly used test case prioritization evaluation metric. The APFD implies how quickly faults are detected by a given ordering of test cases. It is defined as the area under the curve connecting the points (x, y) = (test suite fraction, percentage of faults detected) for a given ordering of test cases. The APFD value ranges from 0 to 1; higher APFD means more effective test case prioritization. We extract the core concept of the APFD as a template and call it APXX (Average Percentage of XX) that implies how quickly XX is satisfied by a given ordering of test cases. Figure~\ref{fig:apxx} visualizes the APXX. To be precise, let $\pi(i)$ be the ordering fraction of the first $i$ test cases for an ordering of $n$ test cases, and let $PXX(\pi(i))$ be the percentage of XX for $\pi(i)$. Note that $PXX(\pi(n)) = 1$ by definition. For an ordering of $n$ test cases, the APXX value as the area under the curve is calculated as follows:
\begin{displaymath}
APXX = \frac{1}{n} \sum_{i=1}^{n} PXX(\pi(i)) - \frac{1}{2n}
\end{displaymath}

Using the APXX template, we define APMK and APMD as follows:
\begin{definition}[APMK and APMD]
For an ordering $\pi$ of a test suite $TS$, the APMK and APMD values are calculated as follows:
\begin{align*}
APMK = \frac{1}{n} \sum_{i=1}^{n} PMK(\pi(i)) - \frac{1}{2n} \\
APMD = \frac{1}{n} \sum_{i=1}^{n} PMD(\pi(i)) - \frac{1}{2n}
\end{align*}
where $n=|TS|$ and $\pi(i)$ is the ordering fraction that contains the first $i$ test cases. 
\end{definition}
In other words, the APMK and APMD imply how quickly mutants are killed and distinguished by a given ordering of tests cases, respectively. As a result, the multi-objective prioritization technique optimizes an ordering of test cases to maximize both APMK and APMD values using a multi-objective optimization algorithm such as NSGA-\rom{2}. As described in Section~\ref{sec:multiObj}, NSGA-\rom{2} returns a set of Pareto optimal orderings, and an additional rule is necessary to choose one of these orderings. MOK selects one of the Pareto optimal orderings that has the highest APMK value. Similarly, MOD selects one of the Pareto optimal orderings that has the highest AMPD value.

\rev{\subsection{Techniques for Comparison}}

\rev{To facilitate our empirical studies, we introduce two simple but widely studied techniques as baselines.}

\paragraph{RND:} We consider random prioritization that randomly prioritizes test cases as a minimum prioritization baseline.

\rev{
\paragraph{SCV:} As an additional control in our studies, we apply the statement-coverage-based test case prioritization.   As explained in Section~\ref{sec:background-prt}, the structural coverage information is widely used surrogate in test case prioritization. We implement the statement-additional that iteratively selects a test case that maximizes the number of additionally covered statements, which is the most effective coverage-based prioritization schemes \cite{henard2016comparing}. If there are multiple test cases that additionally covers the same number of statements, one of them is randomly selected.}

%% file: expr.tex
\section{Experimental Design}\label{sec:expr}

\subsection{Research Questions}
In the experiments, we investigate the following five research questions:
\begin{itemize}
\item RQ1: How do the mutation-based prioritization techniques compare with the random \rev{and coverage-based prioritization in terms of early fault detection?}
\item RQ2: \rev{What is the superior mutation-based prioritization technique in terms of early fault detection?}
\item RQ3: What is the effect of using different weight values in the hybrid (single-objective) test prioritization scheme? 
\item RQ4: How effective are the Pareto front solutions of the multi-objective prioritization scheme?  
\item RQ5: \rev{How much time takes to perform each one of the examined techniques?}
\end{itemize}

RQ1 compares the effectiveness of the mutation-based prioritization techniques with that of the random and coverage-based prioritization. Specifically, we count the number of faults where each of the prioritization techniques is statistically significantly superior, equal, or inferior with the random ordering \rev{and the coverage-based ordering, respectively.} We also measure the effect size of the effectiveness differences of the techniques \rev{with the controls.}

RQ2 compares the effectiveness of the studied techniques among each other with the aim of identifying the best performing technique. Similar to RQ1, we count the number of faults where a technique A is statistically significantly superior, or equal, inferior to another technique B, as well as their exact effectiveness difference.

RQ3 focuses on the hybrid prioritization techniques that uses both the $k$-criterion (i.e., kill) and the $d$-criterion (i.e., distinguish). We examine different weight factors (between kill and distinguish) and see how it impacts the prioritization effectiveness.

RQ4 considers the effectiveness of orderings of test cases in a Pareto front given by the multi-objective test case prioritization techniques. For the multi-objective prioritization, all orderings of test cases in a Pareto front are equally good in terms of the their objectives. However, since the objectives are proxies, the important question is how these orderings perform in terms of the prioritization effectiveness. Thus, for the Pareto front orderings, we investigate the relationship among the prioritization objectives and the prioritization effectiveness.

RQ5 attempts to answer the cost of mutation-based prioritization techniques. One obvious cost of a prioritization technique is the execution time of the technique. We compare the execution times of all the mutation-based prioritization techniques including greedy, hybrid, and multi-objective.

\subsection{Test Subjects and Faults}
For the purposes of the present study, we consider the Java applications in the \texttt{Defects4J} database~\cite{just2014defects4j}. These are all open source software systems and are accompanied by 357 developer-fixed and manually verified real faults. In total, we use the following five applications: JFreeChart (Chart), Closure compiler (Closure), Commons Lang (Lang), Commons Math (Math), and Joda-Time (Time). In \texttt{Defects4J}, each fault is given as an independent fault-fix pair of the program versions.

Out of 357 faults, five faults are excluded because they are not able to give mutation analysis results within a practical time limit (i.e., one-hour per each test case). As a result, we consider the remaining 352 faults, which are summarized in Table \ref{table:subjects}. \rev{Detailed information for each subject fault is available from our webpage at \url{http://se.kaist.ac.kr/donghwan/downloads}.}
	
\begin{table}[]
\centering
\caption{Summary for subject faults, test cases, and mutants. The number of \rev{fault detecting test cases (dT)}, all mutants (aM), killed mutants by the test cases (kM), and distinguished mutants by the test cases (dM) are also presented.}
\label{table:subjects}
\begin{tabular}{lrrrrrr}
\toprule
Program & Faults & Test Cases (sum) & dT (sum) & aM (sum) & kM (sum) & dM (sum)  \\
\midrule
Chart		&	25		& 5,806	& 91		& 21,611	& 8,614		& 1,462 \\
Closure	&	133	& 443,596	& 347	&	109,727	& 82,676	& 34,685 \\
Lang		&	65		& 11,409	& 124	&	81,524	& 63,551	& 5,467 \\
Math		&	106	& 20,661	& 172	&	101,978	& 73,931	& 14,591 \\
Time		&	27		& 72,005	& 76 &	19,996	& 13,665	& 3,838 \\
\midrule
Total		& 352	& 553,477	& 810	& 334,836	& 242,437	& 60,043 \\
\bottomrule
\end{tabular}
\end{table}

\subsection{Test Suites}
For each fault, \texttt{Defects4J} provides ``relevant" JUnit test cases that touch the modified classes between the faulty version and the fixed version. \rev{Test prioritization is performed when testing newly introduced changes. Thus, it is reasonable to use the information about the modified classes, and focus on them instead of the whole program. This is common practice in industry and is performed by retrieving the test cases that have a dependence with the files that were changed \cite{MemonGNDNSM17}. To account for this issue in our experiments, we compose a test suite of relevant test cases for each fault we consider.}

JUnit test cases are Java classes that contain one or more test methods. It leads two different test suite granularity by considering JUnit test cases as the test-class level and the test-method level~\cite{do2006use}. We use the test-method level because it is finer and more informative than the test-class level. In Table \ref{table:subjects}, the column Test Cases (sum) shows the sum of the number of test cases for each fault. For example, there are total 5,806 test cases for the 25 Chart faults. \rev{The column dT (sum) shows the sum of the number of fault detecting test cases for each fault. For example, there are total 91 fault detecting test cases for the 25 Chart faults. Total 553,477 test cases including 810 fault detecting test cases are considered for the 352 subject faults.}

\subsection{Mutants}
We use \texttt{Major}~\cite{just2014major} mutation analysis tool for generating and executing all mutants to the test cases for each fault. It provides a set of commonly used set of mutation operators~\cite{andrews2006using, KintisPPVM17} including the AOR (Arithmetic Operator Replacement), LOR (Logical Operator Replacement), COR (Conditional Operator Replacement), ROR (Relational Operator Replacement), ORU (Operator Replacement Unary), STD (STatement Deletion), and LVR (Literal Value Replacement). We applied all the mutation operators. Since the use of sufficient mutation operators may affect on the experimental results, we will discuss this issue in Section \ref{sec:threats}. 

\rev{We generate mutants out of the fixed (i.e., clean) version of each fault. To perform a controlled experiment, we assume the fixed version is the norm, and perform mutation analysis on it: subsequently, we ``reverse" the fix patch to recreate the fault, and evaluate our prioritization. We will discuss this in Section~\ref{sec:threats} as well.}

We generate mutants only from the modified classes between the fixed version and the faulty version, as we considered only the relevant test cases. In Table \ref{table:subjects}, the column aM (sum), kM (sum), and dM (sum) show the sum of the number of all generated mutants, killed mutants by the test cases, and distinguished mutants by the test suite for each fault, respectively. For example, for the 25 faults in the Chart program, 8,614 mutants and 1,462 mutants among 21,611 mutants are killed and distinguished by the test cases, respectively.

\subsection{Multi-Objective Algorithm Configuration}
For NSGA-\rom{2}, we set the population size as 100. The chosen genetic operators are ones that are widely used for permutation type representation: partially matched crossover, swap mutation, and binary tournament selection~\cite{goldberg1988genetic,epitropakis2015empirical}. The crossover rate is set to 0.9, and the mutation rate is set to 0.2. The maximum fitness evaluation is set to 100,000. \rev{Since finding the best configuration for the mutation-based test case prioritization falls out of the scope of our work, we simply follow the default configuration and parameter values which is commonly used and tuned. Using the default parameter values is a common practice and has been found to be suitable for our context~\cite{kotelyanskii2014parameter}, i.e., search-based testing.}

\subsection{Variables and Measures}
For independent variables, RQ1, RQ2, RQ5 manipulates all the prioritization techniques listed in Table~\ref{table:techniques}, whereas RQ3 and RQ4 focus on the hybrid techniques and the multi-objective techniques, respectively.

For dependent variables, we mainly measure the quality and the cost of the test case prioritization techniques. For the quality of the prioritization, we measure the APFD value for each ordering of test cases. For the cost of the prioritization, we measure the execution time for each ordering of test cases. To provide statistical analysis, we independently generate 100 orderings of test cases for \rev{each of the greedy, hybrid, and control techniques.} For each of the multi-objective techniques, we independently generate 30 orderings of test cases because it takes too long (more than hours for one ordering in the longest case). \rev{All our experiments were performed on the Microsoft Azure Clould Platform using the Ubuntu 16.04 operating system on 8 DS3v2 (4 vcpus, 14 GB memory) virtual machines.}

To compare the effectiveness of two prioritization techniques, we perform statistical hypothesis tests following the guideline provided by Arcuri and Briand \cite{arcuri2014hitchhiker}. \rev{We perform the Mann-Whitney U-test to assess the difference in stochastic order, that is, whether the APFD values in one technique are more likely to be larger than the APFD values in the other technique. Note that the Mann-Whitney U-test is a non-parametric test which makes no assumption about the distribution of the data.} To reduce Type \rom{1} error, the significance level is $\alpha = 0.001$. \rev{We also measure the Vargha and Delaney’s $\hat{A}_{12}$ statistics~\cite{vargha2000critique} to represent the effect size of the effectiveness difference between the compared prioritization techniques. It measures the probability that one technique yields higher APFD values than the other. For example, $\hat{A}_{12}=0.7$ means that one technique outperforms the other in 70\% of the runs.}


For the calculation of APFD values, we use the following equation:
\begin{equation}\label{eq:APFD}
APFD = \frac{1}{n} \sum_{i=1}^{n} PFD(\pi(i)) - \frac{1}{2n}
\end{equation}
where $PFD(\pi(i))$ is the percentage of faults detected by the ordering fraction $\pi(i)$. We should note that there is another commonly used equation provided by Elbaum et al.~\cite{elbaum2002test} as follows:
\begin{equation}\label{eq:oldAPFD}
APFD = 1 - \frac{TF_1+\cdots+TF_n}{nm} + \frac{1}{2n}
\end{equation}
\rev{where $TF_j$ is the first test case position among $n$ test cases which detects the $j$th fault among $m$ faults.} Both (\ref{eq:APFD}) and (\ref{eq:oldAPFD}) give the same APFD value, whereas (\ref{eq:APFD}) uses the percentages of faults detected by test suite fraction and (\ref{eq:oldAPFD}) uses the positions of the first test case that detects each of faults.

\rev{To investigate the relationship between the prioritization effectiveness (i.e., APFD) and objectives (i.e., APMK\footnote{We do not need to additionally investigate the relationship between APMD and APFD because there is a clear inverse relationship between APMK and APMD in Pareto fronts.}), we measure the Pearson linear correlation and Spearman rank correlation between APFD and APMK for the orderings in Pareto fronts. When Pearson (or Spearman) correlation is 1, it means that APFD perfectly linearly (or monotonically) increases as APMK increases for the Pareto optimal orderings. When Pearson (or Spearman) correlation is -1, it means that APFD perfectly linearly (or monotonically) decreases as APMK increases for the Pareto optimal orderings. Consequently, the closer to +1 the correlation is, the more effective MOK is, and the closer to -1 the correlation is, the more effective MOD is.}

%% file: result.tex
\section{Results and Analysis}\label{sec:results}

\subsection{RQ1: Comparison with \rev{Controls}}

Table~\ref{table:withR} records the results for the comparison of the prioritization techniques with the random orderings. For every compared pair (A, B), the column Superiority provides the number of subject faults where the effectiveness of A is statistically superior (+), equal (=), or inferior (-) to B\rev{, based on the Mann-Whitney U-tests with $\alpha=0.001$}. \rev{The column Effect size provides the average $\hat{A}_{12}$ statistics to represent how much one technique outperforms the other in average.} In terms of superior cases, HYB-010 is the best where 86.4\% (304/352) of the subject faults show that the effectiveness of HYB-010 is statistically superior than that of random. In terms of inferior cases, GRD is the best where only 2.27\% (8/352) of the subject faults show that the effectiveness of GRD is statistically inferior than that of random. In terms of effect size, HYB-015 is the best where the \rev{$\hat{A}_{12}$ value is 0.8520. Overall, the mutation-based test case prioritization techniques are statistically superior than or equal to random for 88.9\% of the subject faults.}

\begin{table}
\centering
\caption{Comparison of prioritization effectiveness with random. For every pair (A, B), there are the number of cases where the effectiveness of A is statistically superior (+), equal (=), or inferior (-) to B\rev{, based on the Mann-Whitney U-tests with $\alpha=0.001$}. \rev{The average $\hat{A}_{12}$ value is given to represent the effect size.}}
\resizebox{\textwidth}{!}{
\label{table:withR}
\begin{tabular}{cc | ccc | c || cc | ccc | c}
\toprule
\multicolumn{2}{c|}{Pair} & \multicolumn{3}{c|}{Superiority} & Effect size & \multicolumn{2}{c|}{Pair} & \multicolumn{3}{c|}{Superiority} & Effect size \\
A               & B      & +         & =        & -        & \rev{$\hat{A}_{12}$}  & A               & B      & +         & =        & -        & \rev{$\hat{A}_{12}$}   \\
\midrule
GRD     & RND & 294 & 50 & \textbf{8}  & \rev{0.8269} & HYB-060 & RND & 298   & 37   & 17   & \rev{0.8480} \\
HYB-005 & RND & 298 & 38 & 16 & \rev{0.8447} & HYB-065 & RND & 298   & 36   & 18   & \rev{0.8482} \\
HYB-010 & RND & \textbf{304} & 34 & 14 & \rev{0.8519} & HYB-070 & RND & 299   & 35   & 18   & \rev{0.8474} \\
HYB-015 & RND & 300 & 39 & 13 & \rev{\textbf{0.8520}} & HYB-075 & RND & 301   & 33   & 18   & \rev{0.8475} \\
HYB-020 & RND & 298 & 40 & 14 & \rev{0.8498} & HYB-080 & RND & 299   & 36   & 17   & \rev{0.8473} \\
HYB-025 & RND & 296 & 41 & 15 & \rev{0.8476} & HYB-085 & RND & 300   & 34   & 18   & \rev{0.8484} \\
HYB-030 & RND & 296 & 41 & 15 & \rev{0.8468} & HYB-090 & RND & 297   & 38   & 17   & \rev{0.8488} \\
HYB-035 & RND & 296 & 39 & 39 &\rev{ 0.8475} & HYB-095 & RND & 300   & 35   & 17   & \rev{0.8478} \\
HYB-040 & RND & 295 & 39 & 18 & \rev{0.8476} & GRK     & RND & 275   & 56   & 21   & \rev{0.8188} \\
HYB-045 & RND & 297 & 38 & 17 & \rev{0.8471} & MOK     & RND & 296   & 43   & 13   & \rev{0.8396} \\
HYB-050 & RND & 298 & 36 & 18 & \rev{0.8479} & MOD     & RND & 294   & 47   & 11   & \rev{0.8401} \\ \cline{7-12}
HYB-055 & RND & 298 & 36 & 18 & \rev{0.8476} & Average     & RND & 296.8 & 39.2 & 17.0 & \rev{0.8452} \\
\bottomrule
\end{tabular}}
\end{table}

Table~\ref{table:withR} also shows that hybrid techniques are at least effective as the simple greedy techniques GRD and GRK. Specifically, HYB-095 is more effective than GRK (i.e., HYB-100), even the weight for the GRD is only 0.05. This signifies that it is more effective to consider the $k$-criterion and the $d$-criterion together than to consider the $k$-criterion only.

Interestingly, multi-objective techniques are relatively ineffective than the hybrid techniques. It means that, in comparison with random, multi-objective optimization techniques using the $k$-criterion and the $d$-criterion are less beneficial than merely merging the two greedy techniques.

\begin{table}
\centering
\caption{\rev{Comparison of prioritization effectiveness with coverage-based prioritization. For every pair (A, B), there are the number of cases where the effectiveness of A is statistically superior (+), equal (=), or inferior (-) to B\rev{, based on the Mann-Whitney U-tests with $\alpha=0.001$}. The average $\hat{A}_{12}$ value is given to represent the effect size.}}
\resizebox{\textwidth}{!}{
\label{table:withCV}
\begin{tabular}{cc | ccc | c || cc | ccc | c}
\toprule
\multicolumn{2}{c|}{Pair} & \multicolumn{3}{c|}{Superiority} & Effect size & \multicolumn{2}{c|}{Pair} & \multicolumn{3}{c|}{Superiority} & Effect size \\
A               & B      & +         & =        & -        & $\hat{A}_{12}$  & A               & B      & +         & =        & -        & $\hat{A}_{12}$   \\
\midrule
GRD     & SCV & 238 & 36 & 78 & 0.7012 & HYB-060 & SCV & 266   & 30   & 56   & 0.7813 \\
HYB-005 & SCV & 257 & 29 & 66 & 0.7514 & HYB-065 & SCV & 267   & 29   & 56   & 0.7820 \\
HYB-010 & SCV & 267 & 25 & 60 & 0.7687 & HYB-070 & SCV & 267   & 30   & \textbf{55}   & 0.7822 \\
HYB-015 & SCV & \textbf{268} & 28 & 56 & 0.7791 & HYB-075 & SCV & 267   & 29   & 56   & 0.7820 \\
HYB-020 & SCV & 264 & 32 & 56 & 0.7756 & HYB-080 & SCV & 266   & 31   & \textbf{55}   & 0.7822 \\
HYB-025 & SCV & 262 & 31 & 59 & 0.7736 & HYB-085 & SCV & 266   & 30   & 56   & 0.7836 \\
HYB-030 & SCV & 263 & 31 & 58 & 0.7738 & HYB-090 & SCV & 265   & 32   & \textbf{55}   & \textbf{0.7838} \\
HYB-035 & SCV & 263 & 30 & 59 & 0.7739 & HYB-095 & SCV & 266   & 30   & 56   & 0.7830 \\
HYB-040 & SCV & 264 & 32 & 56 & 0.7772 & GRK     & SCV & 236   & 58   & 58   & 0.7501 \\
HYB-045 & SCV & 267 & 27 & 58 & 0.7793 & MOK     & SCV & 253   & 39   & 60   & 0.7444 \\
HYB-050 & SCV & 267 & 28 & 57 & 0.7806 & MOD     & SCV & 250   & 38   & 64   & 0.7367 \\ \cline{7-12}
HYB-055 & SCV & 267 & 29 & 56 & 0.7809 & Average & SCV & 261.6 & 31.9 & 58.5 & 0.7699 \\
\bottomrule
\end{tabular}}
\end{table}

\rev{Table~\ref{table:withCV} records the results related to the comparison of the prioritization techniques with SCV. The structure of the table is as same as Table~\ref{table:withR}. In terms of superior cases, HYB-015 is the best where 76.1\% (268/352) of the subject faults show that the effectiveness of HYB-015 is statistically superior than that of SCV. In terms of inferior cases, HYB-070, HYB-080, and HYB-090 are the best where 15.6\% (55/352) of the subject faults show that the effectiveness of them are statistically inferior than that of SCV. In terms of effect size, HYB-090 is the best where the $\hat{A}_{12}$ value is 0.7838. Overall, the mutation-based test case prioritization techniques are statistically superior than or equal to the coverage-based prioritization technique at least 77.8\% of the subject faults.}

\begin{framed}
\rev{The mutation-based prioritization is superior to or equal to the random prioritization for 88.9\% of the faults, and is superior to or equal to the coverage-based prioritization for 77.8\% of the faults.}
\end{framed}

\subsection{RQ2: Comparison between the Techniques}\label{sec:allPair}

This section investigates whether there is a superior technique or not among the mutation-based test case prioritization techniques. We only consider GRD, HYB-010, HYB-050, HYB-090, GRK, MOK, and MOD, because there are too many pairs containing all weights for the hybrid techniques. Table~\ref{table:allPair} contains the comparison results for the pair of the techniques. The structure of the table is the same as Table~\ref{table:withR}.

\begin{table}
\centering
\caption{Comparison of prioritization effectiveness of all mutation-based techniques. For every pair (A, B), there are the number of cases where the effectiveness of A is statistically superior (+), equal (=), or inferior (-) to B\rev{, based on the Mann-Whitney U-tests with $\alpha=0.001$}. \rev{The average $\hat{A}_{12}$ value is given to represent the effect size.}}
\resizebox{\textwidth}{!}{
\label{table:allPair}
\begin{tabular}{cc | ccc | c || cc | ccc | c}
\toprule
\multicolumn{2}{c|}{Pair} & \multicolumn{3}{c|}{Superiority} & Effect size & \multicolumn{2}{c|}{Pair} & \multicolumn{3}{c|}{Superiority} & Effect size \\
A               & B      & +         & =        & -        & $\hat{A}_{12}$  & A               & B      & +         & =        & -        & $\hat{A}_{12}$   \\
\midrule
HYB-010 & GRD     & 218 & 74  & 60  & 0.6780 & HYB-090 & HYB-050 & 81 & 226 & 45  & 0.5415 \\
HYB-050 & GRD     & 212 & 62  & 78  & 0.6633 & GRK & HYB-050 & 83 & 188 & 81  & 0.5078 \\
HYB-090 & GRD     & 215 & 55  & 82  & 0.6637 & MOK     & HYB-050 & 40 & 170 & 142 & 0.3913 \\
GRK     & GRD     & 214 & 50  & 88  & 0.6441 & MOD     & HYB-050 & 75 & 113 & 164 & 0.4029 \\
MOK     & GRD     & 122 & 116 & 114 & 0.5049 & GRK     & HYB-090 & 47 & 225 & 80  & 0.4562 \\
MOD     & GRD     & 138 & 127 & 87  & 0.5409 & MOK     & HYB-090 & 38 & 163 & 151 & 0.3830 \\
HYB-050 & HYB-010 & 102 & 169 & 81  & 0.5285 & MOD     & HYB-090 & 79 & 100 & 173 & 0.3951 \\
HYB-090 & HYB-010 & 109 & 163 & 80  & 0.5411 & MOK     & GRK     & 71 & 137 & 144 & 0.4274 \\
GRK & HYB-010 & 108 & 147 & 97  & 0.5200 & MOD     & GRK     & 99 & 84  & 169 & 0.4230 \\
MOK     & HYB-010 & 62  & 139 & 151 & 0.4003 & MOD     & MOK     & 47 & 259 & 46  & 0.5077 \\
MOD     & HYB-010 & 70  & 144 & 138 & 0.4277 &         &         &    &     &     &       \\
\bottomrule
\end{tabular}}
\end{table}

\rev{
Comparing GRK and GRD in Table~\ref{table:allPair}, GRK is more effective at 60.8\% (214/352) faults, whereas GRD is more effective at 25\% (88/352) faults. There is no statistical difference for the remaining 14.2\% (50/352) faults. For all subject faults, the average effect size $\hat{A}_{12}$ is 0.6441, which means that GRK outperforms GRD with the probability of an average 64.41\% of all runs. While GRK is more effective than GRD in general, GRD outperforms GRK for some faults. Section~\ref{sec:discussion} discusses the effectiveness difference between GRK and GRD in more detail.

Interestingly, in comparison to GRD, the average $\hat{A}_{12}$ values of the hybrid techniques are higher than 0.66, while that of the multi-objective techniques are around 0.52. In other words, in comparison to GRD, the hybrid techniques are more effective, while the multi-objective optimization techniques are not. Considering mutant kill and distinguishment together, simple hybrid is more effective than multi-objective optimization in mutation-based test case prioritization.
}

Comparing MOK and MOD, they are equally effective at 73.6\% (259/352) faults, and it is almost the same when MOK is more effective and MOD is more effective for the remaining faults. This implies that orderings of test cases in a Pareto front may have similar prioritization effectiveness. This issue will be investigated in Section~\ref{sec:front}.

Overall, Table~\ref{table:allPair} shows that all pairs have both superior and inferior cases that cannot be ignored. Also, the average $\hat{A}_{12}$ of all pairs are not dramatic. It means that, in terms of mutation-based test case prioritization using the $k$-criterion and the $d$-criterion, there is no single superior technique among greedy, hybrid, and multi-objective.

\begin{framed}
Among greedy, hybrid, and multi-objective strategies using the traditional kill-only mutation adequacy and the diversity-aware mutation adequacy, there is no single superior test case prioritization technique.
\end{framed}

\subsection{RQ3: Effect of Changing Weight between Kill and Distinguish}

\begin{figure}
	\centering
	\includegraphics[width=0.9\textwidth]{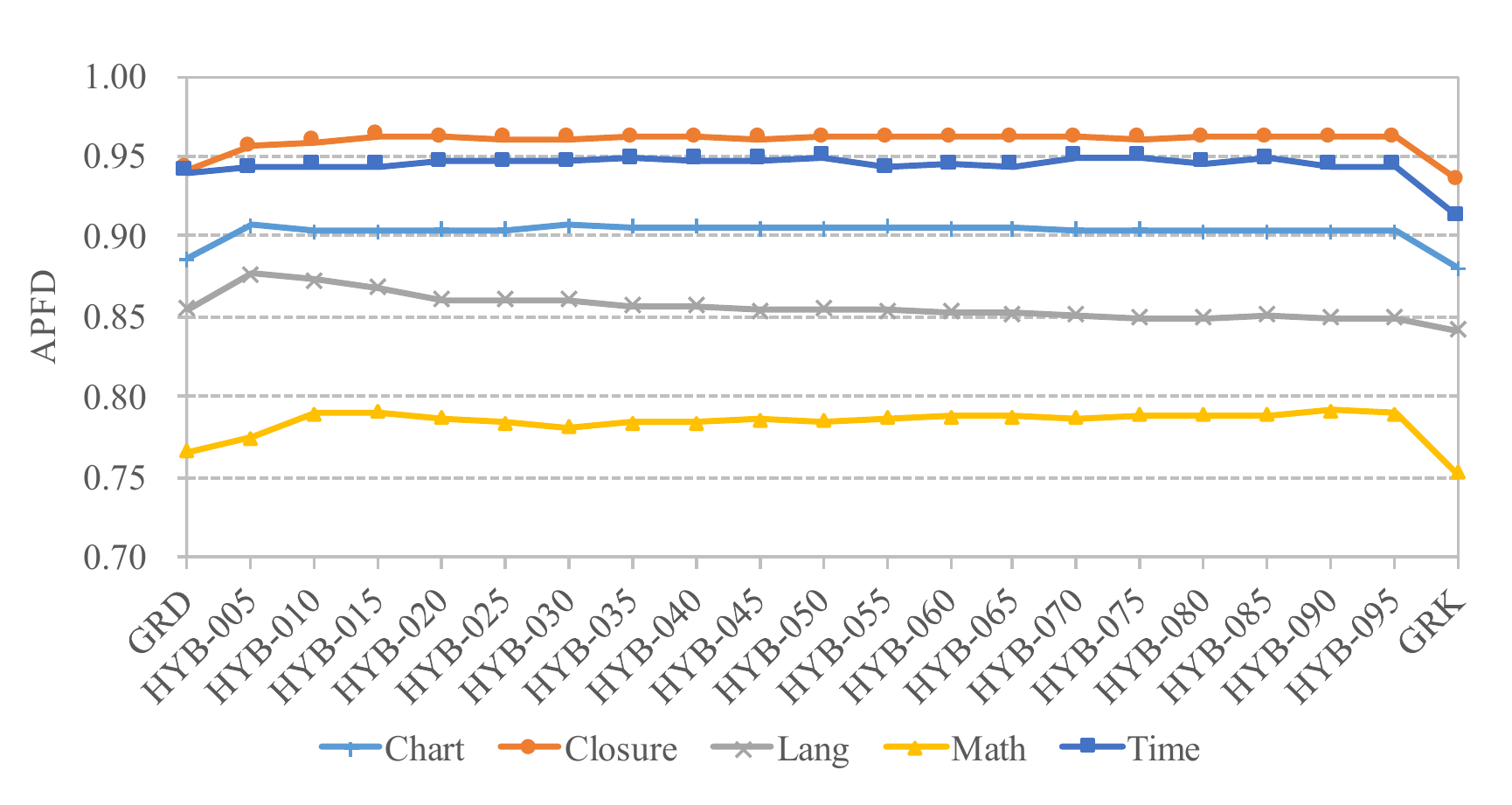}
	\caption{Effect of changing the weight factor in hybrid technique. The effectiveness is maximized when the weight is between 0 and 1 and not on the extreme values 0 or 1.}
	\label{fig:weights}
\end{figure}

To investigate the effect of weight $w$ change on APFD for the HYB-$w$ prioritization techniques, the average APFD is obtained by changing $w$ from 0 to 1 in steps of 0.05. Figure~\ref{fig:weights} shows the results; the x-axis is $w$ and the y-axis is the APFD.

In Figure~\ref{fig:weights}, all the subject programs show the same result: the highest APFD is when $w$ is between 0 and 1 (i.e., neither 0 nor 1). This means that the combination of GRK and GRD has a positive effect on the test case prioritization effectiveness.

There is no single $w$ value showing the highest APFD for all programs. For Chart and Lang, $w=0.05$ shows the highest APFD. For Closure, $w=0.15$ shows the highest APFD. For Math and Time, $w=0.90$ and $w=0.75$ shows the highest APFD, respectively. It means that the best weight $w$ between GRK and GRD depends the program characteristics.

\begin{framed}
The test case prioritization effectiveness is maximized when the weight is between 0 and 1 and not on the extreme values 0 and 1. This means that the combination of GRK and GRD increases the effectiveness. The optimal weight depends on the subject programs.
\end{framed}

\subsection{RQ4: Effectiveness of Orderings in Pareto Fronts}\label{sec:front}
\rev{This section investigates the effectiveness of orderings in Pareto fronts. To do that, we measure the Pearson and Spearman correlation coefficients between APMK (i.e., how quickly mutants are killed) and APFD (i.e., how quickly faults are detected) for the orderings in Pareto fronts.} For 207 among the 352 subject faults (i.e., 58.8\%), the correlation coefficients are undefined because the variance of APFD is zero. It means that, for the 207 fault, all the orderings in a Pareto front are equally good in terms of APFD. This partially explains the fact that MOK and MOD are statistically equally effective at 74.1\% faults as noted in Section~\ref{sec:allPair}. Remaining 145 faults have correlation coefficients ranging from -1 to +1.

\begin{figure}
    \centering
    \subfloat[Pearson correlation]{\label{fig:pearson}\includegraphics[width=0.45\textwidth]{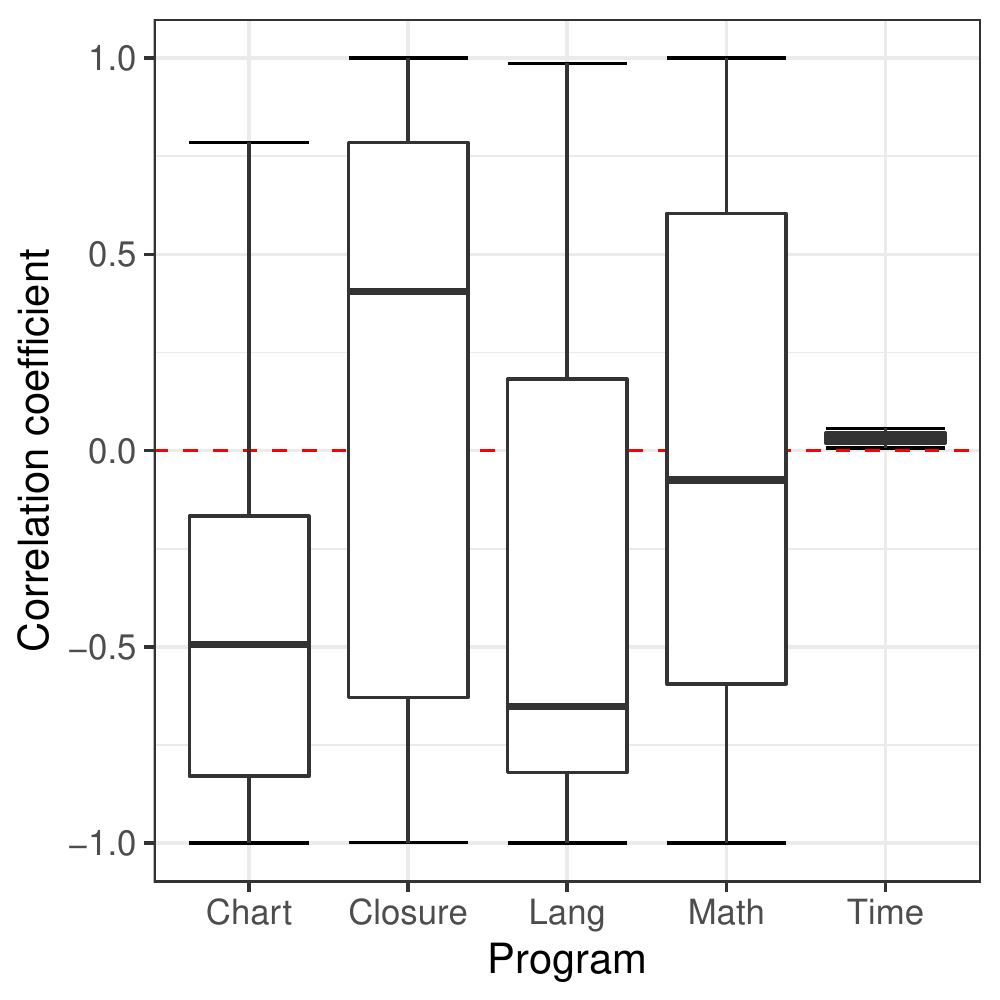}}
    \qquad
    \subfloat[Spearman correlation]{\label{fig:spearman}\includegraphics[width=0.45\textwidth]{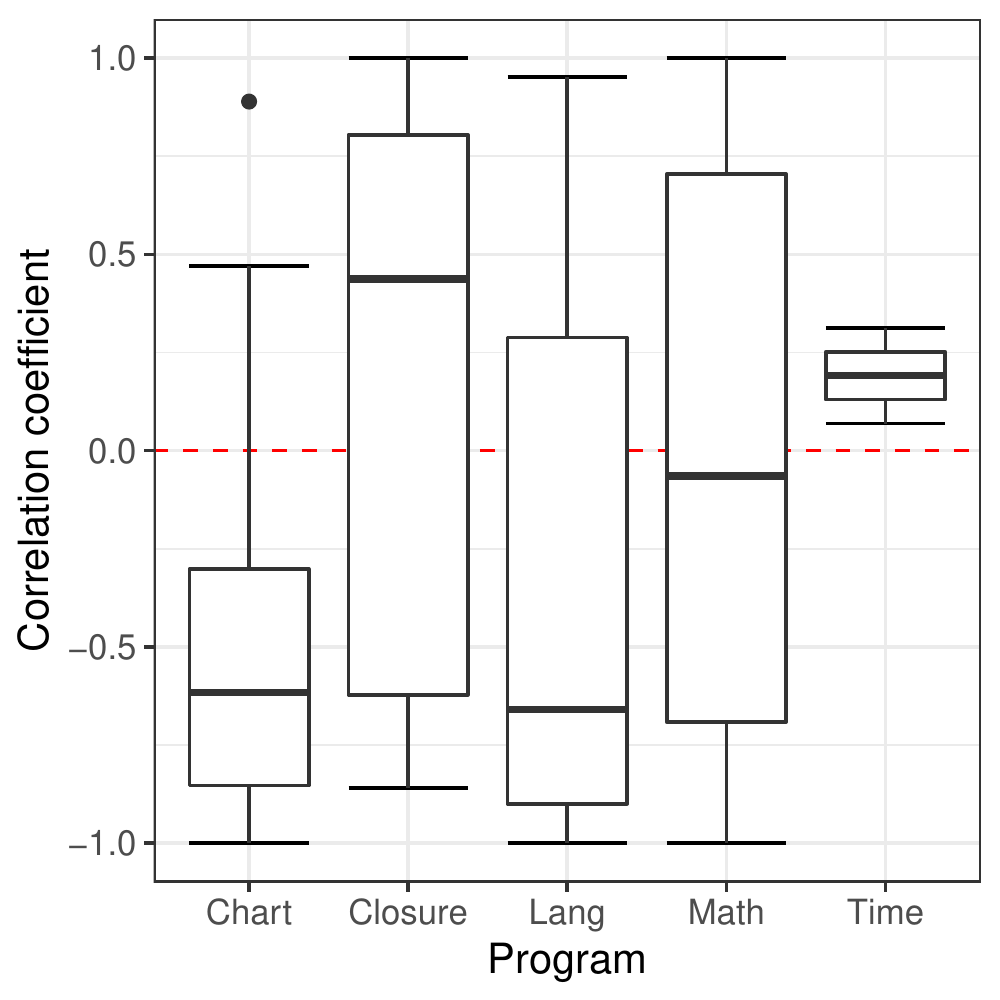}}
    \caption{Correlation coefficients between APMK and APFD for orderings in Pareto fronts. \rev{Each point represents the (a) Pearson or (b) Spearman correlation coefficient of a fault. The bottom and top of the box are the first and third quartiles, and the band inside the box is the median. The top and bottom of the whisker are the highest and lowest datum still within 1.5 IQR of the upper and lower quartile. The distribution of the correlation coefficients widely varies except the faults in the Time program.}}
    \label{fig:paretoSolutions}
\end{figure}

Figure~\ref{fig:paretoSolutions} summarizes the distribution of correlation coefficients of the 145 (=352-207) faults. \rev{Each point represents the Pearson (in Figure~\ref{fig:pearson}) or Spearman (in Figure~\ref{fig:spearman}) correlation coefficient of a fault. The bottom and top of the box are the first and third quartiles, and the band inside the box is the median. The bottom and top of the whiskers are the lowest and highest datum still within 1.5 IQR of the lower and upper quartile. For example, Figure~\ref{fig:pearson} shows that all faults in the Time program have very similar Pearson correlation coefficients, which is nearly zero. This zero Pearson coefficient means that there is no linear correlation between APMK and APFD for the orderings in Pareto fronts.} Except the Time program, Figure~\ref{fig:paretoSolutions} shows that both correlation coefficients are widely distributed from -1 to +1. This implies that there is no superiority between MOK and MOD on average for all programs. The faults in Chart and Lang tend to have correlation coefficients close to -1, whereas the faults in Closure tend to have correlation coefficients close to +1. It implies that MOK is often more effective than MOD for Chart and Lang, whereas MOD is often more effective than MOK for Closure. The faults in Math and Time tend to have zero correlation, which implies that the effectiveness of MOK and MOD is similar.

\begin{framed}
For 58.8\% of the subject faults, the orderings of test cases in Pareto fronts are equally effective in terms of APFD. For the remaining faults, the correlation coefficient between APMK (or APMD) and APFD vary from -1 to +1, depending on the studied faults.
\end{framed}

\subsection{RQ5: Execution Time of the Techniques}

Table~\ref{table:time} shows the average test case prioritization time for each technique. For example, the GRK prioritization technique takes 2253.9 ms to prioritize a test suite on average. There is no time difference between MOK and MOD since both MOK and MOD simply select one orderings of test cases in a Pareto front, and the information needed for selection (i.e., APMK and APMD) is calculated beforehand. 

\begin{table}
\centering
\caption{Execution time for each prioritization technique. The multi-objective techniques require the most execution time, which is approximately 37 minutes. The greedy and hybrid techniques require less than 8 seconds.}
\label{table:time}
\begin{tabular}{lr}
\toprule
Technique & Time (ms) \\
\midrule
RND	& 21.2 	 \\
\rev{SCV} & \rev{150.3} \\
GRK	& 2253.9 	\\
HYB-050	& 7245.2 \\
GRD	& 7651.7	\\
MOK (or MOD)	& 2198981.8 \\
\bottomrule
\end{tabular}
\end{table}

In Table~\ref{table:time}, GRD takes around 3.4 times more time than GRK. This is because the computation for mutant distinguishment (i.e., whether a mutant's d-vector is unique) is harder than the computation for mutant kill (i.e., whether a mutant's d-vector is non-zero). HYB is similar to GRD, because it also requires the computation for mutant distinguishment. While GRD and HYB techniques take more time than GRK, it is within 8 seconds for each prioritization on average. On the other hand, MOK (or MOD) takes far much time; approximately 37 minutes for each prioritization. This is mainly because the number of test cases and mutants are too large to optimize permutations as a whole.

We also investigate the effect of the total number of test cases (i.e., the size of a test suite) and the total number of mutants on the execution time. It turns out that the product of the total number of test cases and the total number of mutants is linearly proportional to the time for all the subject test case prioritization technique. The average Pearson correlation coefficient between the product and the time for all the techniques is 0.930. 

\rev{Note that Table~\ref{table:time} only reports the execution time of the prioritization, not the time for mutation analysis. On average, mutation analysis takes 651.8 seconds per fault. However, there are several test cases that do not give mutation analysis results within one-hour time limit. While such test cases are excluded in our controlled experiment, it can be problematic in practice. Fortunately, mutation analysis for each test case can be easily parallelized. Further, it is possible to prepare the mutation analysis results independently from the future changes and regression testing.}

\begin{framed}
On average, multi-objective techniques requires approximately 37 minutes, whereas greedy and hybrid techniques require less than 8 seconds. The prioritization execution time for all techniques has an exact linear relationship with the product of the number of test cases and mutants.
\end{framed}

\subsection{Discussion}\label{sec:discussion}
\rev{As described in Section~\ref{sec:allPair}, there is no clear winner between GRK (i.e., kill mutants as early as possible) and GRD (i.e., distinguish mutants as early as possible) test prioritization schemes; 60.8\% of the subject faults show that GRK is statistically superior than GRD, whereas 25\% of the subject faults show the opposite result. This is interesting because the $d$-criterion \emph{subsumes} the $k$-criterion as explained in Section~\ref{sec:mutationAdequacy}. Taken together, the $d$-criterion is stronger than the $k$-criterion, whereas the prioritization based on the $d$-criterion is not superior than the prioritization based on the $k$-criterion. To further understand why this happens, we investigate the relationship between kill and distinguish in test case prioritization.

By definition, the mutant kill concerns the difference between the original program and its mutants, whereas the mutant distinguishment concerns the difference among all programs including the original program and mutants. To see how a set of test cases kills and distinguishes mutants and where the fault detecting test cases are, we propose a graphical representation called \emph{Mutant Distinguishment Graph} (MDG). In an MDG, each node represents a set of undistinguished mutants, and a directed edge from a node $n_x$ to another node $n_y$ represents a set of test cases that distinguishes $n_y$ from $n_x$ by killing the mutants in $n_y$ not $n_x$. We call $n_y$ as the child of $n_x$ when there is a directed edge from $n_x$ to $n_y$. To represent where the fault detecting test cases are, we draw the edges with thickness to be proportional to the percentage of fault detecting test cases (among the set of test cases represented by the edge). To avoid zero thickness, we give a default value even if the percentage is zero. There is a special ``root" node that has no incoming edge. In other words, all the remaining nodes are the children of the root. The root node refers the original program and the mutants that are not killed by all the test cases in the MDG. The structure of an MDG varies depending on the given set of test cases and mutants.}

\begin{figure}
	\centering
	\includegraphics[width=0.4\textwidth]{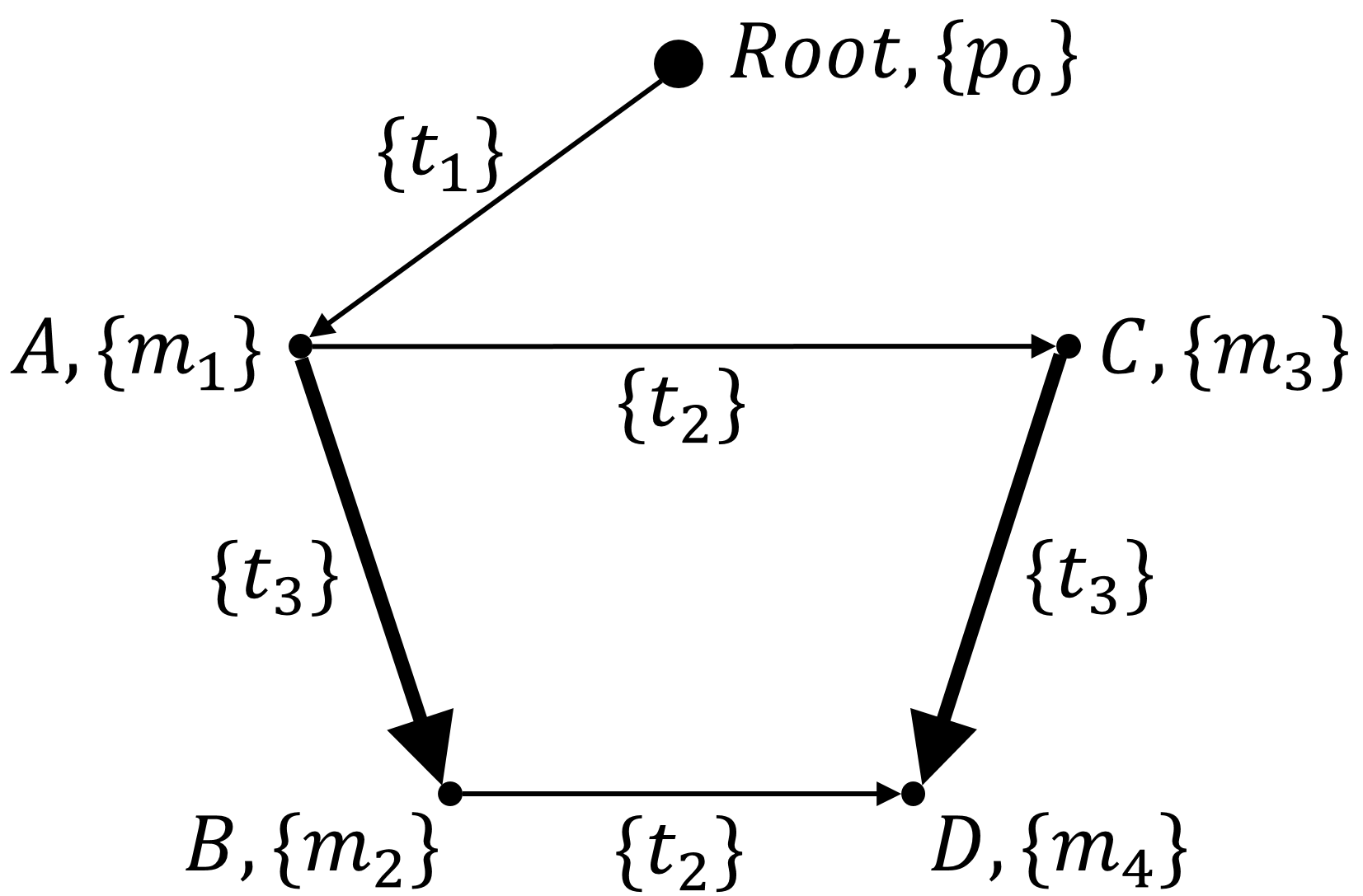}
	\caption{\rev{The MDG corresponding to the example in Figure~\ref{fig:example}}}
	\label{fig:MDG-example}
\end{figure}

\rev{For example, using the four mutants and three test cases given in Figure~\ref{fig:example}, we have the MDG as Figure~\ref{fig:MDG-example}. There are five nodes, $Root$, $A$, $B$, $C$, and $D$, with the five directed edges between the nodes. The $Root$ node at the top only has $p_o$ because all the mutants are killed by the set of test cases $\{t_1,t_2,t_3\}$. The edge from $Root$ to $A$ is labeled with $t_1$. This shows $m_1$ is distinguished from $p_o$ by $t_1$. In other words, $t_1$ kills $m_1$. The edge from $A$ to $B$ is labeled with $t_3$, which shows $m_2$ is distinguished from $m_1$ by $t_3$. We assume $t_3$ is the fault detecting test case as an example, and the edges labeled with $t_3$ are thicker than the others.

Note that there is no direct edge from $Root$ to $B$ labeled with $t_1$ in Figure~\ref{fig:MDG-example}, while $t_1$ also kills $m_2$ as well as $m_1$. This is because the \emph{transitivity} of the mutant distinguishment~\cite{shin:2016theoretical}. For mutants $m_x,m_y,m_z$ and test cases $t_x,t_y$, if $m_y$ is distinguished from $m_x$ by $t_x$ and $m_z$ is distinguished from $m_y$ by $t_y$, then $m_z$ is always distinguished from $m_x$ by both $t_x$ and $t_y$. When it comes to an MDG, such transitivity implies that the edges between a node and its descendants can be omitted without any information loss. As a result, we are able to know that both $t_1$ and $t_3$ kill $m_2$ while there is no direct edge from $Root$ to $B$. 

The transitivity of an MDG provides an interesting property for the test cases in the edges from $Root$. Among all test cases in an MDG, the test cases in the edges from $Root$ are \emph{sufficient} to kill all the mutants except mutants in $Root$. Furthermore, a test case in the edges from $Root$ kills all the mutants in the distinguished node and its descendants. For the example in Figure~\ref{fig:MDG-example}, $\{t_1\}$ (i.e., the set of test cases in the edges from $Root$) is sufficient to kill all mutants, and $t_1$ kills all the mutants in $A$ (i.e., the distinguished node) and $B$, $C$, and $D$ (i.e., the descendants of $A$).

The aforementioned property is an important key to understand the effectiveness difference between GRK and GRD in test case prioritization. For GRK, giving the test cases in the edges from $Root$ high priorities is clearly beneficial to kill all mutants as early as possible. In other words, GRK gives the test cases not in the edges from $Root$ low priorities. If there is no thick edges from $Root$, it means the fault detecting test cases are not in the edges from $Root$, and GRK gives the fault detecting test cases low priorities. In summary, GRK becomes ineffective when there is no thick edge from $Root$. 

GRD tries to distinguish all mutants as early as possible. However, as GRD has to choose among test cases that distinguish mutants (select any edge instead of those from the $Root$) it is likely to give less priority to the test cases in the edges from $Root$. Thus, it is likely to give higher priorities on test cases that distinguish mutants than killing mutants. On the contrary, when fault detecting test cases are triggered by mutatn distinguishement (thick edges are not in the $Root$), we see that GRD becomes effective. In these cases GRD is more likely to outperform GRK.}

\begin{figure}
    \centering
    \subfloat[Closure-18]{\label{fig:Closure-18}\includegraphics[width=0.45\textwidth]{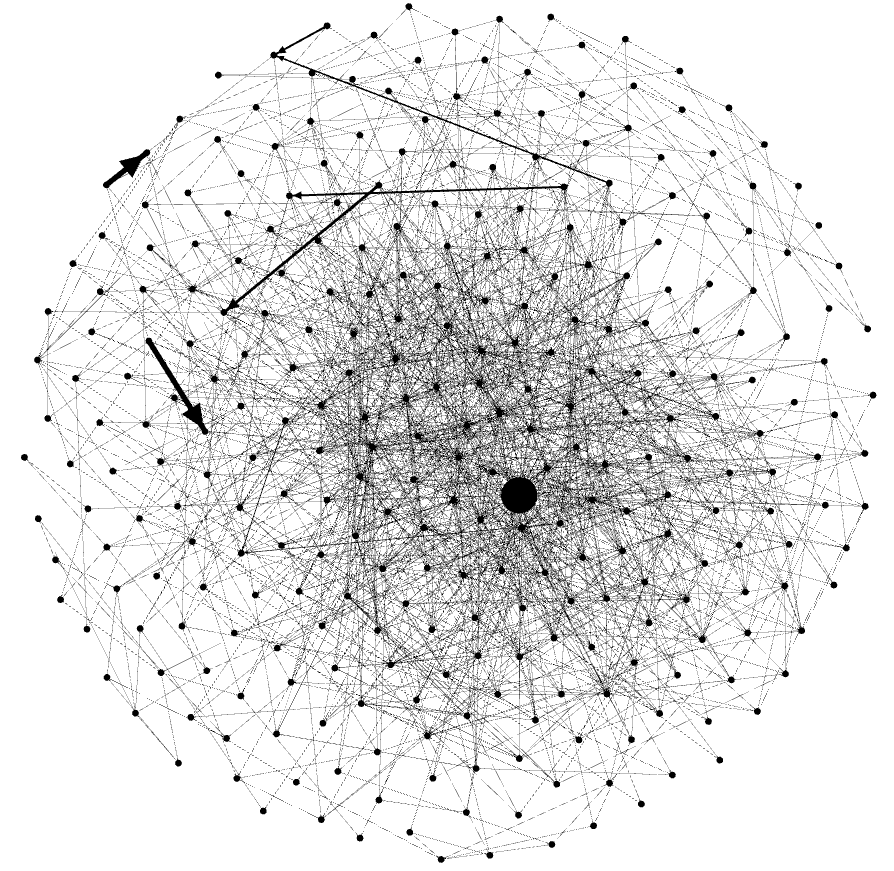}}
    \qquad
    \subfloat[Time-4]{\label{fig:Time-4}\includegraphics[width=0.45\textwidth]{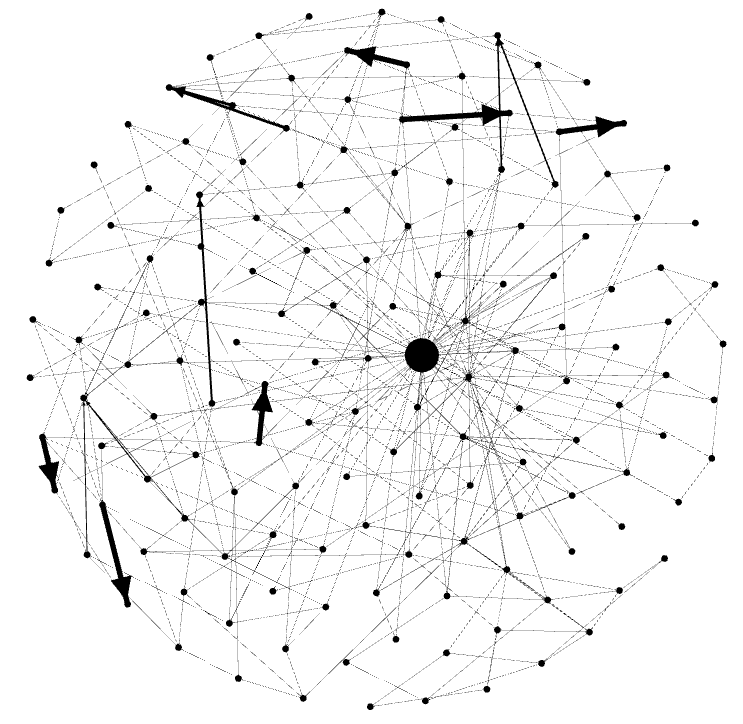}}
    \qquad
    \subfloat[Math-19]{\label{fig:Math-19}\includegraphics[width=0.45\textwidth]{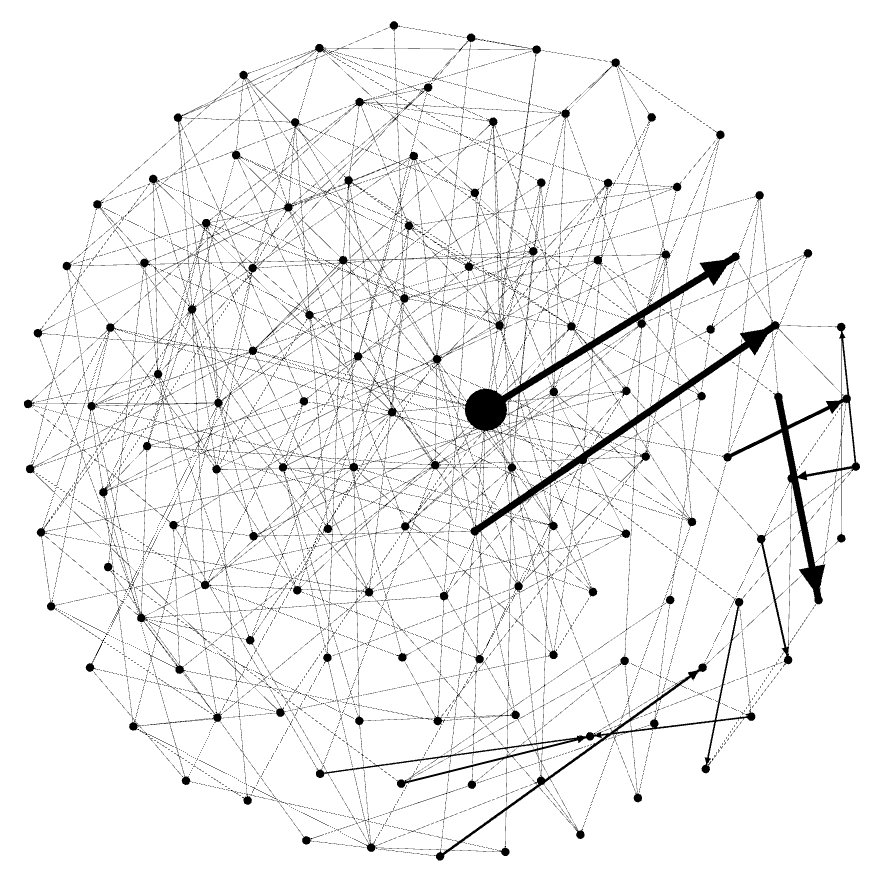}}
    \qquad
    \subfloat[Closure-35]{\label{fig:Closure-35}\includegraphics[width=0.45\textwidth]{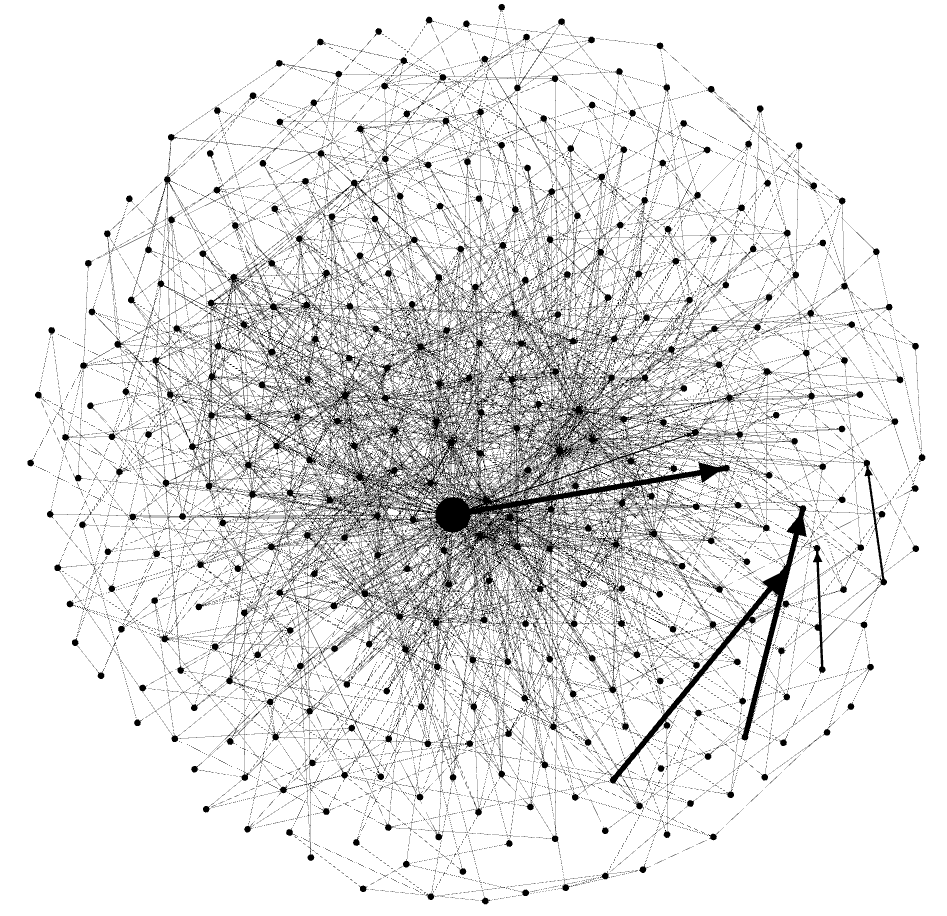}}
    \caption{\rev{MDGs for representative cases. The big node near the center refers to $Root$. When there is no thick edge from $Root$, such as Closure-18 and Time-4, GRK is more likely to be ineffective than GRD. On contrary, when there is a thick edge from $Root$, such as Math-19 and Closure-35, GRK is more likely to be effective than GRD.}}
    \label{fig:MDGs}
\end{figure}

\rev{Figure~\ref{fig:MDGs} shows the four representative MDGs from the 352 subject faults: Figure~\ref{fig:Closure-18} and Figure~\ref{fig:Time-4} show the MDGs for the faults that GRD is much more effective than GRK, whereas Figure~\ref{fig:Math-19} and Figure~\ref{fig:Closure-35} show the faults that GRK is much more effective than GRD. For each graph, the big node near the center refers $Root$. It is clear that there is no thick edge from $Root$ in Figure~\ref{fig:Closure-18} and Figure~\ref{fig:Time-4}, whereas there is a thick edge from $Root$ in Figure~\ref{fig:Math-19} and Figure~\ref{fig:Closure-35}. 

Unfortunately, we cannot have an MDG without the fault detection information. Since we do not know which test case detects faults in prioritization time, we cannot use an MDG to predict which of GRK and GRD will be more effective than the other. Thus, MDGs are useful in explaining the strengths and weakness of the mutation-based test case prioritization schemes. To improve test case prioritization, we need some form of prediction related the properties of the fault detecting test cases. If such an a prediction is made then we can use MDGs to also support test case prioritization.}


We should note that an MDG is similar to the Mutant Subsumption Graph (MSG) suggested by Kurtz et al.~\cite{kurtz2014mutant}, as the mutant distinguishment is closely related to the mutant subsumption as discussed by Shin and Bae~\cite{shin:2016theoretical}. The main difference between an MDG and an MSG is that an MDG additionally contains the information of the fault detecting test cases in the thickness of edges.

\subsection{Threats to Validity}\label{sec:threats}
There are several threats to validity for our experimental results. One threat is due to the subject programs and faults that we use. These might not be representative of other programs and faults. While this threat is common to any empirical study and can only be addressed by making multiple and context-related studies, we tried to mitigate it by using a large set of real faults. Thus, we used all the faults of \texttt{Defects4J}, which is an independently constructed dataset built to support controlled experimental results. 

Our results are also to some extent dependent on the configuration of NSGA-\rom{2}. \rev{Parameter turning is an important and challenging problem in evolutionary algorithms~\cite{eiben1999parameter}. However, we feel that such a tuning will not impact much our results as the default configurations perform well in our context \cite{kotelyanskii2014parameter}.}

\rev{The mutation analysis tool \texttt{Major}~\cite{just2014major} and the mutation operators we use form another source of threats for our study. This is because different types of mutants may result in different behaviour and influence our results. Therefore, the use of another mutation testing tool employing a different set of operators, like the \texttt{PIT}~\cite{coles2016pit}, may result in different findings. However, we do not consider this threat as vital as our main contribution lies in the relative comparison of the mutation-based test prioritization techniques and not on their optimal performance. Moreover, we expect that using different mutant sets will have a similar influence on all the prioritization techniques we study since all of them rely on the same set of mutants. Nonetheless, according to a recent study by Kintis \etal \cite{KintisPPVM17}, the fault detection capabilities of \texttt{PIT} are considerably lower than those of \texttt{Major}. The same study reports that another version of the tool, named \texttt{PIT\_RV} (i.e., the research version of \texttt{PIT}), is 5\% more effective at detecting faults than \texttt{Major}. Therefore, we believe that this 5\% difference from \texttt{PIT\_RV} cannot make a major difference on the results we report.}



\rev{Other threats come from the order of the fixed and faulty versions. While the fixed version comes after the faulty version in the repository timeline, we assume the fixed version as the clean version that previously passes all regression test cases and the faulty version as the change-introduced version that should be tested by the regression test cases. Such reverse order is used to perform our controlled experiment. Still more studies are needed in order to investigate the differences between the results of the controlled experiments and actual practice.}

The APFD metric used for representing the effectiveness of test case prioritization has some limitations. It does not account for the severity of faults and test case execution cost. \rev{Since \texttt{Defects4J} provides many single-fault program versions instead of one multiple-faults program version, we do not need to concern the severity of each fault. To overcome the limitation related to the test case execution cost, we additionally measure the APFD$_c$ values~\cite{elbaum2001incorporating} which account for the execution times of individual test cases. The results show that the average difference between the APFD and APFD$_c$ values for each test case ordering is almost negligible (i.e., 3.169e-04). This is because the execution times of test cases in a test suite are almost equivalent for all the subject test suites. As a result, we keep use the APFD metric for representing the results.}

\rev{To allow reproducibility of the results presented in this paper, all the prioritization results and the implementation of the mutation-based test case prioritization techniques are available from our web page at \url{http://se.kaist.ac.kr/donghwan/downloads}.}

%% file: relatedWork.tex
\section{Related Work}\label{sec:related}

Since the diversity-aware mutation adequacy criterion (i.e., the $d$-criterion) has been recently proposed by Shin et al.~\cite{shin2016diversity}, there is no previous study for the diversity-aware mutation adequacy in test case prioritization. However, the $d$-criterion is experimentally evaluated in test suite selection, compared to the traditional kill-only mutation adequacy criterion (i.e., the $k$-criterion). The results on 45 real faults in \texttt{Defects4J} show that the $d$-criterion increases the fault detection effectiveness of adequate test suites in comparison with the $k$-criterion, whereas the $d$-criterion requires more test cases to be adequate than the $k$-criterion.

In test case prioritization, the traditional mutation adequacy criterion is already investigated by Rothermel et al.~\cite{rothermel2001prioritizing}. They investigate the effectiveness of several greedy prioritization techniques using branches, statements, and mutants, respectively. The branch and statement techniques prioritizes test cases according to the number of branches and statements covered by each test case, respectively. They find that there is no single best technique. However, on average across the programs, the mutant-based technique performs most effectively. Later, Elbaum et al.~\cite{elbaum2002test} extend the empirical study of Rothermel et al. by including function-level coarser granularity techniques in comparison with the statement-level fine granularity techniques. The empirical results on eight C programs listed in Siemens benchmarks~\cite{hutchins1994experiments} show that the coarser granularity decreases the effectiveness of test case prioritization in general. 

For the total greedy approach and the additional greedy approach in test case prioritization, Li et al.~\cite{li2007search} report that the additional approach significantly outperforms the total approach. They also study meta-heuristic algorithms for test case prioritization, whereas the prioritization effectiveness difference between the performance of meta-heuristic and that of additional greedy is not significant. Zhang et al.~\cite{zhang2013bridging} also focus on the total and additional approaches. They develop a unified approach with the total and additional at two extreme instances. The unified model yields a spectrum of genetic approaches ranging between the total and additional approaches depending on a control parameter. The empirical results on four Java programs show that selecting a proper parameter increases the prioritization effectiveness compared to the simple total and additional approaches. However, the additional approach is almost effective as the parametrized approach in all programs.

In multi-objective test case prioritization, Epitropakis et al.~\cite{epitropakis2015empirical} present an empirical study of the effectiveness of multi-objective test case prioritization. They mainly investigate two different multi-objective evolutionary algorithms, NSGA-\rom{2} and Two Archive Evolutionary Algorithm (TAEA)~\cite{deb2000fast}, for the objectives including statement coverage and fault detection history. The results show that the multi-objective prioritization techniques are superior to greedy techniques that target each of the objectives of the multi-objective technique.

Perhaps the work that is the closest to ours is that of Lou et al.~\cite{lou2015mutation}, which studies mutation-based prioritization within software evolution. The study concerns two prioritization schemes; one based on the number of mutants killed and one based on the distribution of the killed mutants. Their results show that ordering tests by the number of mutants killed performs best. This approach is similar to our greedy one. While there are similarities between our and Lou et al. studies, our is based on real faults (while theirs is based on mutant-faults) and we consider the distinguish method with multiple heuristics, while they do not.

Regarding diversity-based test prioritization, there are several studies working mainly in a black-box manner. Henard et al.~\cite{henard2014bypassing} suggest diversity-aware metric based on the concept of Combinatorial Interaction Testing. This method performs test prioritization by ordering tests according to the dissimilarity of the combinations of the test input parameters. Feldt et al.~\cite{feldt2016test} suggest using a compression utilities to support test prioritization. The techniques measures the dissimilarity distance of test suites using the concept of Normalized Compression Distance. More recently, Hennard et al.~\cite{henard2016comparing} compare these techniques with other coverage-based test prioritization and find that they are of similar power despite that they do not use any dynamic information from the tested systems.

%% file: conclusion.tex
\section{Conclusion}\label{sec:conclusion}
In this paper, we investigate test case prioritization guided by mutants. Based on the recently defined diversity-aware mutation adequacy criterion, we present the new prioritization objective that distinguishing all mutants as early as possible. We evaluate the effectiveness of mutation-based prioritization techniques by considering the new objective as well as the existing objective, which is killing all mutants as early as possible. Based on these two objectives, we investigate greedy, hybrid, and multi-objective prioritization strategies using 352 real faults and 553,477 developer-written test cases.

\rev{Our results show that the mutation-based prioritization is more than or equally effective than the random prioritization and the coverage-based prioritization for at least 88.9\% and 77.8\% of the faults, respectively. Among the greedy, hybrid, and multi-objective optimization strategies using the kill-only and diversity-aware mutation adequacy criteria, there is no single superior test case prioritization technique. Interestingly, while there is no superiority between the kill-only mutation and the diversity-aware mutation adequacy criteria, their combined use improves the effectiveness of the prioritization. For the multi-objective optimization, the effectiveness of orderings in Pareto fronts does not have steady correlation with the prioritization objectives. The prioritization execution time for the multi-objective techniques requires approximately 37 minutes, while the greedy and hybrid techniques require less than 8 seconds.

There are several implications from the results. For example, both distinguishing and killing mutants as early as possible are more effective than covering statements as early as possible. To detect faults as early as possible, the mutation-based prioritization is more beneficial than the coverage-based one. Interestingly, by considering the two mutation-based objectives in one greedy hybrid approach performs best. The same combination can be done by using a multi-objective optimization but unfortunately it does not provide any important benefits and requires far more time to prioritize than the hybrid. 

More research is needed in order to develop a single test case prioritization technique that is clearly superior to killing and distinguishing mutants. To support such attempts, we provide a graphical model called Mutant Distinguishment Graph (MDG), which visualizes how mutants are killed and distinguished by a test suite with respect to the fault detecting test cases. This way we demonstrate the reasons why simply killing mutants as early as possible is not always effective.}